\documentclass[11pt]{article}
\usepackage{makecell}

\usepackage[preprint]{acl}

\usepackage{times}
\usepackage{latexsym}
\usepackage{url}
\usepackage[T1]{fontenc}

\usepackage[utf8]{inputenc}

\usepackage{microtype}

\usepackage{inconsolata}

\usepackage{graphicx}

\usepackage{tcolorbox}
\tcbuselibrary{breakable}
\usepackage{multirow}
\usepackage{booktabs}
\usepackage{fancyhdr}
\usepackage[table]{xcolor}

\definecolor{rank_1}{rgb}{0.45, 0.65, 0.90}
\definecolor{rank_2}{rgb}{0.58, 0.74, 0.93}
\definecolor{rank_3}{rgb}{0.70, 0.82, 0.96}
\definecolor{rank_4}{rgb}{0.80, 0.88, 0.98}
\definecolor{rank_5}{rgb}{0.89, 0.93, 0.99}
\definecolor{rank_6}{rgb}{0.95, 0.97, 1.00}

\definecolor{c_0_10}   {rgb}{0.88, 0.42, 0.42} 
\definecolor{c_10_20}  {rgb}{0.95, 0.55, 0.55}
\definecolor{c_20_30}  {rgb}{0.97, 0.62, 0.62}
\definecolor{c_30_40}  {rgb}{1.00, 0.70, 0.70}
\definecolor{c_40_45}  {rgb}{1.00, 0.80, 0.80}
\definecolor{c_45_50}  {rgb}{1.00, 0.89, 0.89} 
\definecolor{c_50_55}  {rgb}{0.90, 1.00, 0.90} 
\definecolor{c_55_60}  {rgb}{0.78, 0.975, 0.78}
\definecolor{c_60_70}  {rgb}{0.65, 0.95, 0.65}
\definecolor{c_70_80}  {rgb}{0.55, 0.90, 0.55}
\definecolor{c_80_90}  {rgb}{0.45, 0.85, 0.45}
\definecolor{c_90_100} {rgb}{0.35, 0.75, 0.35} 

\title{A Global View of How LLMs Value Human Time}

\author{Yifan He \\
  University of Antwerp, Belgium \\
  \texttt{yifan.he@uantwerpen.be} \\
  \textbf{Sofie Goethals} \\
  University of Antwerp, Belgium \\
  \texttt{sofie.goethals@uantwerpen.be} \\
  \And Manon Reusens \\
  University of Antwerp, Belgium \\
  \texttt{manon.reusens@uantwerpen.be} \\
  \textbf{David Martens} \\
  University of Antwerp, Belgium \\
  \texttt{david.martens@uantwerpen.be} \\}

\begin{document}

\maketitle
\begin{abstract}
Large language models (LLMs) increasingly act as personal assistants; however, little is known about how they capture national-level preferences on a global scale. Given that ``time'' is a universal and representative concept, we examine LLMs' time preferences by varying only users' country and measuring the compensation LLMs deem appropriate for a 60-minute appointment delay. We then compare LLMs' responses with GDP per capita on Purchasing Power Parity (GDP\_PPP), Mean Income, and a literature-based index (Patience Score). Our results across 193 countries and 5 LLMs demonstrate that people's time across the world is evaluated in noticeably different ways among LLMs, as well as varying levels of robustness under prompt variations. 
LLMs consistently recommend the highest compensation for Europe, but mostly the lowest for Africa. Based on GDP\_PPP, LLMs underestimate the value of time for users living in Asia and South America. These findings indicate that geographical patterns are embedded in LLMs and may translate in different treatment in agentic settings.

\end{abstract}

\section{Introduction}

People rely on LLMs in daily life, asking for emotional support, general advice, and carrying out autonomous tasks~\cite{zao-sanders2026ai}. Many of these tasks can require some kind of trade-off, where a positive outcome should be put in perspective against a negative one. For example, waiting a bit longer before your pizza is being delivered while paying a lower price. When outsourcing these decisions to LLMs, it is important to know how these choices are made, as these can vary greatly, depending on the model used, language of the prompt, users' gender, and so on~\citep{10.1145/3799434}. Users from different countries may reasonably expect LLMs to personalise their recommendations based on relevant local circumstances, while ensuring that they are not disadvantaged. 
However, LLMs often respond from a European-centric, Western-related and English-dominant perspective~\citep{10.1093/pnasnexus/pgae346,  goethals-rhue-2025-one, durmus2024measuringrepresentationsubjectiveglobal, Buyl2026}. Most papers only examine a subset of countries and cultures, potentially overlooking important global consequences~\citep{10.1162/COLI.a.14}. 




\begin{figure}
    \centering
    \includegraphics[width=\linewidth]{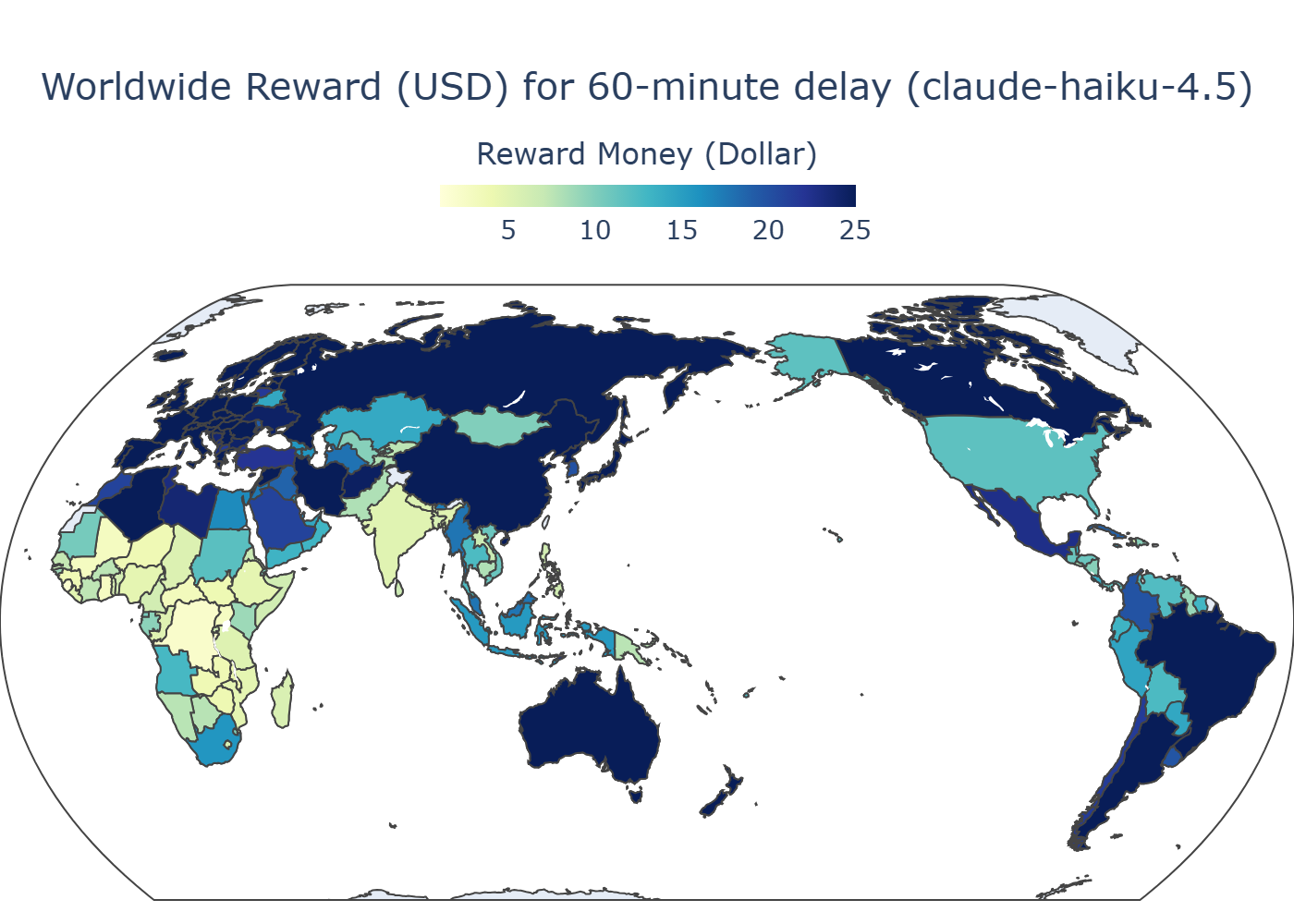}
    \caption{Example of the derived compensation for a 60-minute delay in Claude-Haiku-4.5. This illustrates the differences in estimated time value across the globe.}
    \label{fig:intro}
\end{figure} 

\begin{figure*}[h!]
\centering
\includegraphics[width=0.8\linewidth]{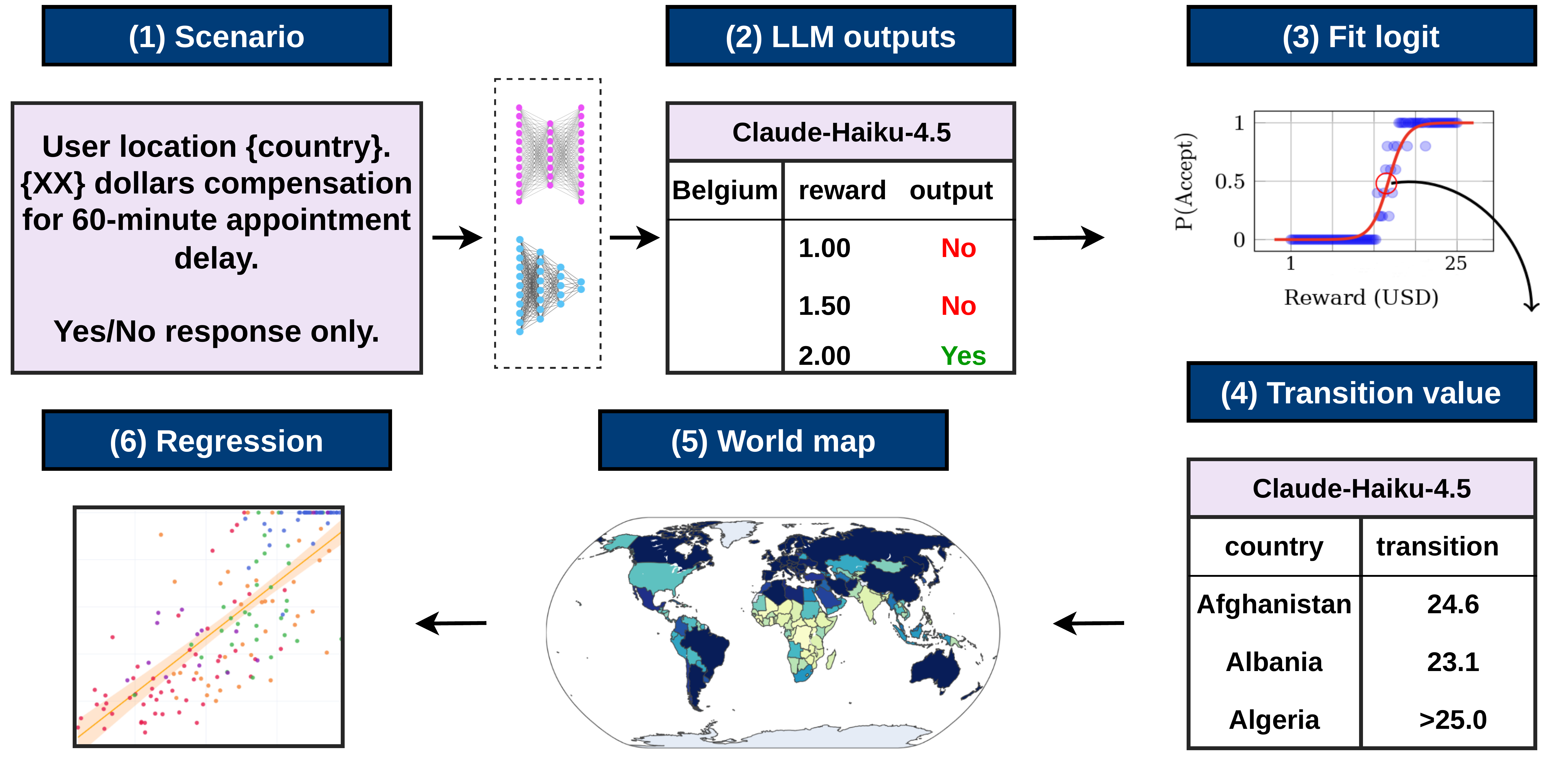}
  \caption {Overview of the workflow. LLMs decide whether to accept a series of compensations for additional waiting. For each country, a transition value is determined at which LLMs flip answers using logistic regression. These numbers are visualised on world maps and compared against multiple indicators.}
  \label{schematic}
\end{figure*}
 
In this study, we offer a global perspective by examining responses and deviations across all countries from the United Nations Member States list. LLMs are asked to determine monetary compensation for a user's additional waiting time. 
We use the methodology introduced in \citet{10.1145/3799434} to calculate the compensation per country, represented by when LLMs change answers from ``No'' to ``Yes'' given a series of offers.
An example of the result from Claude-Haiku-4.5\footnote{The code for the experiments can be found on Anonymous GitHub: \url{https://github.com/Anthyf/llm_decisions_globally}.} is shown in Figure~\ref{fig:intro}. This figure shows how waiting time is evaluated differently across the globe by the model. 


We make the following contributions. (i) We offer a systematic, global-scale evaluation on the trade-off between monetary compensation and waiting time across different LLMs.
(ii) We analyse our results against economic indicators and a human-derived patience score to compare the LLM-provided answers.
(iii) We uncover systematic geographic patterns across five LLMs and discuss the potential impact of these patterns. 


\section{Related Work}




\paragraph{Human Time Preferences} Social scientists use time-discounting questions as a proxy for people's valuation of time, generating national patience scores to measure cross-country differences \citep{10.1093/qje/qjy013, 10.1371/journal.pone.0245692,Ruggeri2022}. Additionally, research focuses on how much people are willing to pay to save travel time in transportation field~\citep{Whittington_Cook_2019}. 
In economics, the value of time is often conceptualized through its opportunity cost, as time not spent working represents the loss of potential earnings~\citep{NBERw11601}.
This is widely measured by the wage rate (before- or after-tax)~\citep{WARDMAN201693, Whittington_Cook_2019} or GDP~\citep{BINSUWADAN2023126}. 


\paragraph{LLMs Time Preferences} Similarly, multiple works quantify LLMs' time preferences through time-discounting questions~\citep{chen2023emergence,liu2025largelanguagemodelsassume,SUH2026116274, fish2026econevalsbenchmarkslitmustests,10.1287/mksc.2023.0306}. \citet{mazyaki2025temporalpreferenceslanguagemodels} examine whether geographic information in prompts influences temporal preferences in language models for three countries.~\citet{YAN2026101245} assess whether LLMs approximate human valuations of travel time, finding strong alignment between the two.
~\citet{10.1145/3799434} examine the trade-off between waiting time and financial reward, revealing heterogeneity in LLMs’ time preferences. However, a study examining LLMs’ time preferences on a global scale is currently lacking. 

\section{Methodology}

\begin{table*}[htbp]
\centering
\small
\scalebox{0.85}{\begin{tabular}{llcccccc}
\toprule
& Models & Africa (54) & Asia (46)& Europe (44) & N. America (23)& Oceania (14)& S. America (12)\\
\midrule
\multirow{5}{*}{\shortstack{\textbf{Compensation}\\ \textbf{(US\$)}}}
&claude-haiku-4.5 & \cellcolor{rank_6}13.1$\pm$10.0 & \cellcolor{rank_3}16.1$\pm$8.3 & \cellcolor{rank_1}20.7$\pm$6.8 & \cellcolor{rank_4}15.3$\pm$9.0 & \cellcolor{rank_5}15.1$\pm$9.0 & \cellcolor{rank_2}16.8$\pm$8.2 \\
&deepseek-v4-pro & \cellcolor{rank_6}2.3$\pm$0.8 & \cellcolor{rank_4}3.7$\pm$0.9 & \cellcolor{rank_1}6.6$\pm$1.2 & \cellcolor{rank_3}3.8$\pm$0.9 & \cellcolor{rank_2}4.3$\pm$1.1 & \cellcolor{rank_5}3.0$\pm$0.8 \\
&glm-5.2 & \cellcolor{rank_6}1.1$\pm$0.0 & \cellcolor{rank_3}2.7$\pm$0.1 & \cellcolor{rank_1}6.3$\pm$0.8 & \cellcolor{rank_2}3.2$\pm$0.4 & \cellcolor{rank_4}2.6$\pm$0.4 & \cellcolor{rank_5}1.6$\pm$0.3 \\
&gpt-5.4-mini & \cellcolor{rank_6}4.6$\pm$4.1 & \cellcolor{rank_5}5.5$\pm$5.5 & \cellcolor{rank_1}8.8$\pm$9.0 & \cellcolor{rank_2}6.5$\pm$6.4 & \cellcolor{rank_4}5.6$\pm$5.5 & \cellcolor{rank_2}6.5$\pm$6.0 \\
&mistral-large-3 & \cellcolor{rank_4}16.5$\pm$7.5 & \cellcolor{rank_5}16.2$\pm$7.7 & \cellcolor{rank_1}18.5$\pm$6.0 & \cellcolor{rank_3}17.5$\pm$6.6 & \cellcolor{rank_2}18.2$\pm$6.2 & \cellcolor{rank_6}15.4$\pm$8.4 \\
\midrule
& \textbf{Mean} & \textbf{7.5} &\textbf{ 8.8 }& \textbf{12.2 }& \textbf{9.3} & \textbf{9.2} & \textbf{8.7} \\
\bottomrule
\end{tabular}}

\footnotesize{
\textbf{Note.} Mistral-Large-3 assigns all 193 countries a compensation value of \$25 under Prompt 3, yielding a flat fitted line with a $R^2=-\infty$, which distorts the statistic result. We only use the results from Prompts 1 and 2 in this case, marked by $^{\dagger}$. Numbers between brackets represent the number of countries. 
}

\caption{Accepted compensation (US\$) per continent across three prompt variations. }
\label{continent_combined}
\end{table*}
Figure~\ref{schematic} shows an overview of the methodology. 

\paragraph{Experimental Setup.} We assess the accepted financial compensation by LLMs for a one-hour delay in an appointment. We experiment with 193 different countries\footnote{These are the countries of the United Nations Member States list: \url{https://www.un.org/en/about-us/member-states}.} on a range of financial rewards (from USD \$1 to \$25), with 49 evenly spaced increments. We select five models ensuring representativeness for different continents:  Claude-Haiku-4.5, DeepSeek-v4-pro, GLM-5.2, GPT-5.4-Mini, and Mistral-Large-3. 
Experiments are run five times at temperature 0. Implementation details are in Appendix~\ref{setup}. Robustness tests use two paraphrased prompt versions, provided in Appendix~\ref{prompts}.

After the collection of responses, we fit a logistic regression for each country, which outputs a compensation value corresponding to the point at which the LLM shifts its decision from reject to accept~\cite{10.1145/3799434}. Transition values below \$1 or above \$25 is
converted to \$1 or \$25, respectively. These values are visualised on world maps, as shown in Figure~\ref{schematic}.

\paragraph{Economic Indicators and Patience Score.} 

We compare the gathered monetary values against three indicators. First, \textbf{GDP per capita at Purchasing Power Parity (GDP\_PPP)} captures average economic output per person while accounting for cross-country differences in price levels. 
It is also used by \citet{WARDMAN201693} to explain the value of time.
Second, \textbf{Mean Income}, measured as ``Mean income or consumption per day'' from the World Bank Platform~\citep{worldbank2026pip}, captures average daily household per capita income or consumption expenditure.
 Prior research links willingness to save time to income~\citep{WARDMAN2015200, MACKIE20143}.
 Third, \textbf{Patience Score} provides a behavioral measure from human studies and serves as a proxy for time value. We use the country-level scores reported by \citet{10.1371/journal.pone.0245692}. Further details  are in Appendix~\ref{app:econ}.

We run separate linear regressions between LLM-derived time values and each indicator, using ($R^2$) to measure the strength of the association. For GDP\_PPP, we also examine the \textit{residuals}, the deviations between  observed and fitted time values. The residual sign indicates whether the LLM  under- or overestimates a country's value of time compared with the GDP\_PPP-based prediction.

\section{Results}

\subsection{LLMs Time Valuation}

We aggregate accepted compensation by continent across all three prompt variations, and report the means and standard deviations in Table~\ref{continent_combined}. First, we observe substantial differences across LLMs in how they value people's time. By continent, Claude-Haiku-4.5 and Mistral-Large-3 are relatively more generous in their recommendations, whereas the other three models accept compensation mostly below \$8. The differences between countries can be noticeable. Figure~\ref{fig:intro} shows an example from Claude-Haiku-4.5, with substantial differences in accepted compensation ranging from \$1 to the upper bound of \$25. The models also differ in the consistency of their responses across prompt variations. GLM-5.2 and DeepSeek-V4-Pro exhibit greater stability, while the other models show sensitivity to prompt paraphrasing. 
Mistral-Large-3, for example, produces divergent compensation estimates between \$9 to \$18 across continents under Prompt 1 and Prompt 2, but assigns a uniform value of \$25 to all 193 countries under Prompt 3.
 Appendix~\ref{othermodels} provides world maps for all LLMs under Prompt 1, and Appendix~\ref{robustness} provides more detailed results from the three prompt variations.


\begin{table*}[htbp]
\centering
\small
\scalebox{0.9}{\begin{tabular}{l|c|c|c}
\toprule
Models & $R^2$: GDP\_PPP (179) &  $R^2$: Mean Income (113)&  $R^2$: Patience (111) \\
\midrule
claude-haiku-4.5 & $0.261 \pm 0.285$ & $0.264 \pm 0.245$ & $0.075 \pm 0.071$ \\
deepseek-v4-pro & $0.642 \pm 0.046$ & $0.816 \pm 0.014$ & $0.297 \pm 0.022$ \\
glm-5.2 & $0.543 \pm 0.034$  & $0.839 \pm 0.008$  & $0.277 \pm 0.021$ \\
gpt-5.4-mini & $0.289 \pm 0.240$ & $0.392 \pm 0.243$ & $0.160 \pm 0.102$  \\
mistral-large-3 & $0.091^{\dagger} \pm 0.035^{\dagger}$ & $0.237^{\dagger} \pm 0.013^{\dagger}$ & $0.067^{\dagger} \pm 0.030^{\dagger}$  \\
\bottomrule
\end{tabular}}

\footnotesize{\textbf{Note.} Mistral-Large-3 is marked by $^{\dagger}$ as in Table~\ref{continent_combined}. Numbers between brackets represent the number of countries.}

\caption{Linear Regression Results with three indicators: GDP\_PPP, Mean Income, and Patience Score.}
\label{table3}
\end{table*}

\begin{table*}[htbp]
\centering
\small
\scalebox{0.85}{\begin{tabular}{llcccccc}
\toprule
& Models & Africa (54) & Asia (46)& Europe (44) & N. America (23)& Oceania (14)& S. America (12)\\
\midrule
\multirow{5}{*}{\shortstack{\textbf{Overvalued (\%)}\\\textbf{to GDP\_PPP}}}
&claude-haiku-4.5 & \cellcolor{c_60_70}62.2$\pm$31.6 & \cellcolor{c_30_40}39.2$\pm$25.0 & \cellcolor{c_50_55}52.8$\pm$37.0 & \cellcolor{c_40_45}44.4$\pm$36.1 & \cellcolor{c_60_70}61.9$\pm$25.1 & \cellcolor{c_50_55}51.5$\pm$18.9 \\
&deepseek-v4-pro & \cellcolor{c_50_55}53.8$\pm$3.3 & \cellcolor{c_20_30}25.0$\pm$4.3 & \cellcolor{c_50_55}51.2$\pm$2.4 & \cellcolor{c_30_40}39.7$\pm$12.0 & \cellcolor{c_80_90}83.3$\pm$8.2 & \cellcolor{c_0_10}0.0$\pm$0.0 \\
&glm-5.2 & \cellcolor{c_50_55}53.2$\pm$2.9 & \cellcolor{c_20_30}30.0$\pm$6.6 & \cellcolor{c_55_60}56.9$\pm$1.4 & \cellcolor{c_30_40}34.9$\pm$9.9 & \cellcolor{c_60_70}69.0$\pm$14.9 & \cellcolor{c_0_10}0.0$\pm$0.0 \\
&gpt-5.4-mini & \cellcolor{c_40_45}44.2$\pm$3.3 & \cellcolor{c_10_20}19.2$\pm$15.3 & \cellcolor{c_55_60}56.1$\pm$16.9 & \cellcolor{c_30_40}33.3$\pm$16.5 & \cellcolor{c_40_45}40.5$\pm$4.1 & \cellcolor{c_50_55}54.5$\pm$15.7 \\
&mistral-large-3 & \cellcolor{c_50_55} $51.0^{\dagger}\pm1.3^{\dagger}$ & \cellcolor{c_20_30}$27.5^{\dagger}\pm3.5^{\dagger}$ & \cellcolor{c_45_50}$45.1^{\dagger}\pm1.7^{\dagger}$ & \cellcolor{c_50_55}$52.4^{\dagger}\pm13.4^{\dagger}$ & \cellcolor{c_60_70}$67.9^{\dagger}\pm15.2^{\dagger}$ & \cellcolor{c_10_20}$13.6^{\dagger}\pm6.4^{\dagger}$ \\

\midrule
&\textbf{Mean} & \cellcolor{c_50_55}\textbf{52.9} & \cellcolor{c_20_30}\textbf{28.2} & \cellcolor{c_50_55}\textbf{52.4} & \cellcolor{c_40_45}\textbf{40.9} & \cellcolor{c_60_70}\textbf{64.5} & \cellcolor{c_20_30}\textbf{23.9} \\

\bottomrule
\end{tabular}}

\footnotesize{
\textbf{Note.} Mistral-Large-3 is marked by $^{\dagger}$ as in Table~\ref{continent_combined}. Numbers between brackets represent the number of countries. 
}

\caption{The percentage of overvalued countries by
continent across three prompt variations.}
\label{continent_overvalue}
\end{table*}

Several geographical patterns are consistent across all LLMs. European countries consistently get the highest compensation across all five LLMs, followed by North America and Oceania; whereas African countries receive the lowest compensation. However, models show varying assessments on most continents (except for Europe), as reflected by the different shades of blue. For example, Claude-Haiku-4.5 and GPT-5.4-Mini assign the second-highest compensation values to South America, whereas Mistral-Large-3 assigns it the lowest. This broadly aligns with common perceptions of economic contexts, with European countries often viewed as relatively more developed, while regions such as South America and Asia are generally perceived as more varied in their levels of development and economic conditions~\citep{binstock2013influences,THORNTON20121053}.

\subsection{Association Analysis with All Indicators}

Table~\ref{table3} summarises the linear regression results from all three indicators. Among the five LLMs, Mistral-Large-3's valuations are the least explained, as reflected in the lowest $R^2$ across all indicators. In contrast, DeepSeek-V4-Pro and GLM-5.2 show consistently stronger associations with these underlying factors.
Cross-indicator comparisons should be interpreted with caution due to differences in country coverage, but mean income generally explains LLM responses best. DeepSeek-V4-Pro and GLM-5.2 achieve $R^2>0.8$ with low deviations. As time value can be interpreted as an opportunity cost linked to earnings~\citep{NBERw11601, Whittington_Cook_2019}, these models may reflect this relationship through training on related topics. 
GDP\_PPP also explains responses reasonably well, particularly for DeepSeek-V4-Pro and GLM-5.2 ($R^2>0.5$)\footnote{Examples of Linear Regressions from GDP\_PPP are shown in Appendix~\ref{scatterall}. }. However, as a macroeconomic measure that includes components such as government consumption and investment, it does not directly capture individual income, potentially explaining its lower explanatory power. 
Patience score explains little of the variation in LLM responses, with all $R^2$
 values below 0.3. As a composite behavioral measure influenced by cultural, social, and institutional factors, it may capture dimensions not reflected in LLM responses.


\subsection{Overvaluation Rates Relative to GDP\_PPP}

We analyse country-level residuals and calculate, for each continent, the overvaluation rate, which is the percentage of countries whose responses exceed the values fitted by GDP\_PPP, shown in Table~\ref{continent_overvalue}.\footnote{We use GDP\_PPP as an indicator as it covers the most countries while providing moderate explanatory power. We also provide the results that fit three indicators on the same set of countries, as shown in Appendix~\ref{regressionsame}. }
Differences in overvaluation rates across continents suggest systematic regional variation in LLM-generated compensation estimates relative to GDP\_PPP. All five LLMs show lower overvaluation rates in Asia and South America, indicating more frequent undervaluation of time, followed by North America. Europe and Africa show more balanced rates, with greater variation across models. Oceania has the highest overvaluation rate, with an average of 64.5\% of countries receiving compensation above GDP\_PPP-based expectations. Overall, these findings suggest systematic regional patterns in how LLMs value time.



\section{Discussion and Conclusion}
This paper examines how LLMs evaluate the value of people's waiting time from a global perspective. Our results reveal heterogeneity across LLMs in both time valuation and response robustness. When comparing LLMs' responses against established indicators, we do not find strong correlation patterns between model responses and these metrics for all models. Although we do not claim that any of these indicators form a ground truth, and it is reasonable for LLMs to include other socioeconomic and demographic factors into consideration, we still find substantial differences between the LLM responses. These should be transparently communicated to users. 

Hence, our results call for caution when integrating LLMs into daily life as autonomous tools, particularly where their outputs could shape individual decisions. Future work could incorporate multiple indicators, factors such as delay tolerance, and denominate compensation in local currencies to better reflect real-world contexts.


\section*{Limitations}
Our study has several limitations.

\textbf{Indicators and analysis.} First, the choice and interpretation of indicators require caution. For example, GDP\_PPP is a complex macroeconomic measure, and there is no need for LLM responses to strictly reflect it. We fit separate linear regressions with a single variable each time, which might limit the explanatory power ($R^2$). All three linear regressions are also fitted on different set of countries, which mitigates the reliability of the comparisons; nevertheless, the patterns remain unchanged after we check on the same set of countries, shown in Table~\ref{table4} in the Appendix~\ref{regressionsame}. Additionally, we report overvaluation rates only as the percentage of countries per continent. Because overvaluation and undervaluation rates sum to 100\%, a 50\% rate, in our assumption, is regarded as relatively fair for cross-continental comparison, but this captures the occurrence rather than the magnitude of residuals. Examining absolute residuals could give a fuller picture of how closely LLM valuations correlate with GDP\_PPP. Meanwhile, the systematic geographic patterns claimed in this paper that are represented by overvaluation percentages are relative merely to GDP\_PPP, which could be compromised due to the possible divergent results based on other indicators.  
Additionally, perceptions of a 60-minute delay may vary across countries and cultures, meaning the same delay may impose different levels of inconvenience. For example, cultures with stronger norms of punctuality may perceive this delay as more serious than cultures in which delays are more commonly accepted.

\textbf{Outliers.} Special cases can compromise the reliability of these percentages. Under Prompt 3, Claude-Haiku-4.5 assigns \$25 to most countries, so most African countries with low GDP\_PPP are easily classed as overvalued (98.1\%). This outlier inflates the average and produces large standard deviations in Table~\ref{continent_combined}.

\textbf{Study design.} 
Capping compensation at [1,25] may distort country- and continent-level estimates. Additionally, although we fix model checkpoints and providers, accessing LLMs through APIs may still affect reproducibility, as providers can change quantization, serving infrastructure, or the underlying model without notice. Lastly, all offers are denominated in US dollars. Future research could better reflect real-world contexts by expressing the compensation in each country's local currency. 





\section*{Ethical Considerations}
This study reveals that different LLMs yield varying results regarding compensation for a one-hour delay across different countries. On one hand, variations in local wage levels, costs of living, and the actual costs incurred due to delays make a certain degree of variation reasonable; on the other hand, assigning different values to human time could be perceived as unfair. There is no consensus on which approach constitutes real ``fairness''. We do not take a specific stance on which behaviour is more appropriate; rather, we use these metrics such as GDP\_PPP or income levels solely as reference, aiming to explain model behaviour. Meanwhile, LLMs also exhibit substantial differences in their robustness regarding prompt wording. These findings warn researchers and engineers when deploying LLMs as autonomous tools. We hope this work encourages further global-wise evaluations and sparks broad discussion regarding appropriate behaviour. The study did not involve human participants, annotators, or personal data. Country-level indicator data are sourced from the World Bank and used exclusively for research purposes.


\bibliography{bib}

@article{10.1145/3799434,
author = {Cedro, Mateusz and Ichmoukhamedov, Timour and Goethals, Sofie and He, Yifan and Hinns, James and Martens, David},
title = {Cash or Comfort? How LLMs Value Your Inconvenience},
year = {2026},
issue_date = {September 2026},
publisher = {Association for Computing Machinery},
address = {New York, NY, USA},
volume = {69},
number = {9},
issn = {0001-0782},
url = {https://doi.org/10.1145/3799434},
doi = {10.1145/3799434},
journal = {Commun. ACM},
month = sep,
pages = {58–66},
numpages = {9}
}

@article{zao-sanders2026ai,
  author  = {Zao-Sanders, Marc},
  title   = {How People Are Really Using AI in 2026},
  journal = {Harvard Business Review},
  year    = {2026},
  month   = jun,
  day     = {1},
  url     = {https://hbr.org/2026/06/how-people-are-really-using-ai-in-2026}
}

@article{10.1371/journal.pone.0245692,
    doi = {10.1371/journal.pone.0245692},
    author = {Rieger, Marc Oliver AND Wang, Mei AND Hens, Thorsten},
    journal = {PLOS ONE},
    publisher = {Public Library of Science},
    title = {Universal time preference},
    year = {2021},
    month = {02},
    volume = {16},
    url = {https://doi.org/10.1371/journal.pone.0245692},
    pages = {1-15},
    number = {2},

}

@article{Ruggeri2022,
  author  = {Ruggeri, Kai and Panin, Amma and Vdovic, Milica and Ve{\'c}kalov, Bojana and Abdul-Salaam, Nazeer and others},
  title   = {The globalizability of temporal discounting},
  journal = {Nature Human Behaviour},
  year    = {2022},
  volume  = {6},
  pages   = {1386--1397},
  doi     = {10.1038/s41562-022-01392-w},
  url     = {https://www.nature.com/articles/s41562-022-01392-w}
}

@article{10.1093/qje/qjy013,
    author = {Falk, Armin and Becker, Anke and Dohmen, Thomas and Enke, Benjamin and Huffman, David and Sunde, Uwe},
    title = {Global Evidence on Economic Preferences},
    journal = {The Quarterly Journal of Economics},
    volume = {133},
    number = {4},
    pages = {1645-1692},
    year = {2018},
    month = {11},
    issn = {0033-5533},
    doi = {10.1093/qje/qjy013},
    url = {https://doi.org/10.1093/qje/qjy013},
    eprint = {https://academic.oup.com/qje/article-pdf/133/4/1645/28835875/qjy013.pdf},
}

@article{WARDMAN201693,
title = {Values of travel time in Europe: Review and meta-analysis},
journal = {Transportation Research Part A: Policy and Practice},
volume = {94},
pages = {93-111},
year = {2016},
issn = {0965-8564},
doi = {https://doi.org/10.1016/j.tra.2016.08.019},
url = {https://www.sciencedirect.com/science/article/pii/S0965856415302810},
author = {Mark Wardman and V. Phani K. Chintakayala and Gerard {de Jong}}
}

@article{Whittington_Cook_2019, title={Valuing Changes in Time Use in Low- and Middle-Income Countries}, volume={10}, DOI={10.1017/bca.2018.21}, number={S1}, journal={Journal of Benefit-Cost Analysis}, author={Whittington, Dale and Cook, Joseph}, year={2019}, pages={51–72}}

@article{BINSUWADAN2023126,
title = {The income elasticity of the value of travel time savings: A meta-analysis},
journal = {Transport Policy},
volume = {136},
pages = {126-136},
year = {2023},
issn = {0967-070X},
doi = {https://doi.org/10.1016/j.tranpol.2023.03.013},
url = {https://www.sciencedirect.com/science/article/pii/S0967070X23000720},
author = {Jawaher Binsuwadan and Mark Wardman and Gerard {de Jong} and Richard Batley and Phill Wheat}
}

@article{SHIRES2009315,
title = {An international meta-analysis of values of travel time savings},
journal = {Evaluation and Program Planning},
volume = {32},
number = {4},
pages = {315-325},
year = {2009},
note = {Evaluating the Impact of Transport Projects: Lessons for Other Disciplines},
issn = {0149-7189},
doi = {https://doi.org/10.1016/j.evalprogplan.2009.06.010},
url = {https://www.sciencedirect.com/science/article/pii/S0149718909000548},
author = {J.D. Shires and G.C. {de Jong}}
}

@article{SUH2026116274,
title = {Time and risk preferences among generative AI systems},
journal = {Journal of Business Research},
volume = {217},
pages = {116274},
year = {2026},
issn = {0148-2963},
doi = {https://doi.org/10.1016/j.jbusres.2026.116274},
url = {https://www.sciencedirect.com/science/article/pii/S0148296326003097},
author = {Wangsuk Suh and Stephen A. Atlas}
}

@misc{fish2026econevalsbenchmarkslitmustests,
      title={EconEvals: Benchmarks and Litmus Tests for Economic Decision-Making by LLM Agents}, 
      author={Sara Fish and Julia Shephard and Minkai Li and Ran I. Shorrer and Yannai A. Gonczarowski},
      year={2026},
      eprint={2503.18825},
      archivePrefix={arXiv},
      primaryClass={cs.AI},
      url={https://arxiv.org/abs/2503.18825}, 
}

@article{10.1287/mksc.2023.0306,
author = {Goli, Ali and Singh, Amandeep},
title = {Frontiers: Can Large Language Models Capture Human Preferences?},
year = {2024},
issue_date = {July-August 2024},
publisher = {INFORMS},
address = {Linthicum, MD, USA},
volume = {43},
number = {4},
issn = {1526-548X},
url = {https://doi.org/10.1287/mksc.2023.0306},
doi = {10.1287/mksc.2023.0306},
journal = {Marketing Science},
month = jul,
pages = {709–722},
numpages = {14}
}

@article{10.1093/pnasnexus/pgae346,
    author = {Tao, Yan and Viberg, Olga and Baker, Ryan S and Kizilcec, René F},
    title = {Cultural bias and cultural alignment of large language models},
    journal = {PNAS Nexus},
    volume = {3},
    number = {9},
    pages = {pgae346},
    year = {2024},
    month = {09},
    issn = {2752-6542},
    doi = {10.1093/pnasnexus/pgae346},
    url = {https://doi.org/10.1093/pnasnexus/pgae346},
    eprint = {https://academic.oup.com/pnasnexus/article-pdf/3/9/pgae346/59151559/pgae346.pdf},
}

@misc{durmus2024measuringrepresentationsubjectiveglobal,
      title={Towards Measuring the Representation of Subjective Global Opinions in Language Models}, 
      author={Esin Durmus and Karina Nguyen and Thomas I. Liao and Nicholas Schiefer and Amanda Askell and Anton Bakhtin and Carol Chen and Zac Hatfield-Dodds and Danny Hernandez and Nicholas Joseph and Liane Lovitt and Sam McCandlish and Orowa Sikder and Alex Tamkin and Janel Thamkul and Jared Kaplan and Jack Clark and Deep Ganguli},
      year={2024},
      eprint={2306.16388},
      archivePrefix={arXiv},
      primaryClass={cs.CL},
      url={https://arxiv.org/abs/2306.16388}, 
}

@inproceedings{goethals-rhue-2025-one,
    title = "One world, one opinion? The superstar effect in {LLM} responses",
    author = "Goethals, Sofie  and
      Rhue, Lauren",
    editor = "Prabhakaran, Vinodkumar  and
      Dev, Sunipa  and
      Benotti, Luciana  and
      Hershcovich, Daniel  and
      Cao, Yong  and
      Zhou, Li  and
      Cabello, Laura  and
      Adebara, Ife",
    booktitle = "Proceedings of the 3rd Workshop on Cross-Cultural Considerations in NLP (C3NLP 2025)",
    month = may,
    year = "2025",
    address = "Albuquerque, New Mexico",
    publisher = "Association for Computational Linguistics",
    url = "https://aclanthology.org/2025.c3nlp-1.8/",
    doi = "10.18653/v1/2025.c3nlp-1.8",
    pages = "89--107",
    ISBN = "979-8-89176-237-4"
}

@article{10.1162/COLI.a.14,
    author = {Pawar, Siddhesh and Park, Junyeong and Jin, Jiho and Arora, Arnav and Myung, Junho and Yadav, Srishti and Haznitrama, Faiz Ghifari and Song, Inhwa and Oh, Alice and Augenstein, Isabelle},
    title = {Survey of Cultural Awareness in Language Models: Text and Beyond},
    journal = {Computational Linguistics},
    volume = {51},
    number = {3},
    pages = {907-1004},
    year = {2025},
    month = {09},
    issn = {0891-2017},
    doi = {10.1162/COLI.a.14},
    url = {https://doi.org/10.1162/COLI.a.14},
    eprint = {https://direct.mit.edu/coli/article-pdf/51/3/907/2523159/coli.a.14.pdf},
}

@article{chen2023emergence,
  author  = {Chen, Yiting and Liu, Tracy Xiao and Shan, You and Zhong, Songfa},
  title   = {The emergence of economic rationality of {GPT}},
  journal = {Proceedings of the National Academy of Sciences},
  year    = {2023},
  volume  = {120},
  number  = {51},
  pages   = {e2316205120},
  doi     = {10.1073/pnas.2316205120},
  pmid    = {38085780},
  url     = {https://doi.org/10.1073/pnas.2316205120}
}

@misc{mazyaki2025temporalpreferenceslanguagemodels,
      title={Temporal Preferences in Language Models for Long-Horizon Assistance}, 
      author={Ali Mazyaki and Mohammad Naghizadeh and Samaneh Ranjkhah Zonouzaghi and Hossein Setareh},
      year={2025},
      eprint={2509.09704},
      archivePrefix={arXiv},
      primaryClass={cs.CL},
      url={https://arxiv.org/abs/2509.09704}, 
}

@misc{liu2025largelanguagemodelsassume,
      title={Large Language Models Assume People are More Rational than We Really are}, 
      author={Ryan Liu and Jiayi Geng and Joshua C. Peterson and Ilia Sucholutsky and Thomas L. Griffiths},
      year={2025},
      eprint={2406.17055},
      archivePrefix={arXiv},
      primaryClass={cs.CL},
      url={https://arxiv.org/abs/2406.17055}, 
}

@article{YAN2026101245,
title = {Valuing time in silicon: Can large language models replicate human value of travel time},
journal = {Travel Behaviour and Society},
volume = {44},
pages = {101245},
year = {2026},
issn = {2214-367X},
doi = {https://doi.org/10.1016/j.tbs.2026.101245},
url = {https://www.sciencedirect.com/science/article/pii/S2214367X26000141},
author = {Yingnan Yan and Tianming Liu and Yafeng Yin}
}

@dataset{worldbank2026pip,
  author       = {{World Bank}},
  year         = {2026},
  title        = {Poverty and Inequality Platform},
  version      = {20260922\_2021 and 20260922\_2017},
  type         = {Data set},
  publisher    = {World Bank Group},
  url          = {https://pip.worldbank.org/},
  urldate      = {2026-10-05}
}

@article{WARDMAN2015200,
title = {How should business travel time savings be valued?},
journal = {Economics of Transportation},
volume = {4},
number = {4},
pages = {200-214},
year = {2015},
issn = {2212-0122},
doi = {https://doi.org/10.1016/j.ecotra.2015.08.003},
url = {https://www.sciencedirect.com/science/article/pii/S2212012215000416},
author = {Mark Wardman and Richard Batley and James Laird and Peter Mackie and John Bates}
}

@article{MACKIE20143,
title = {Transport appraisal revisited},
journal = {Research in Transportation Economics},
volume = {47},
pages = {3-18},
year = {2014},
note = {Appraisal in Transport},
issn = {0739-8859},
doi = {https://doi.org/10.1016/j.retrec.2014.09.013},
url = {https://www.sciencedirect.com/science/article/pii/S0739885914000547},
author = {Peter Mackie and Tom Worsley and Jonas Eliasson}
}

@techreport{NBERw11601,
 title = "Lifecycle Prices and Production",
 author = "Aguiar, Mark and Hurst, Erik",
 institution = "National Bureau of Economic Research",
 type = "Working Paper",
 series = "Working Paper Series",
 number = "11601",
 year = "2005",
 month = "September",
 doi = {10.3386/w11601},
 URL = "http://www.nber.org/papers/w11601",
}

@article{Buyl2026,
  author  = {Buyl, Maarten and Rogiers, Alexander and Noels, Sander and Bied, Guillaume and Dominguez-Catena, Iris and Heiter, Edith and Johary, Iman and Mara, Alexandru-Cristian and Romero, Raphaël and Lijffijt, Jefrey and De Bie, Tijl},
  title   = {Large language models reflect the ideology of their creators},
  journal = {npj Artificial Intelligence},
  year    = {2026},
  volume  = {2},
  number  = {1},
  pages   = {7},
  doi     = {10.1038/s44387-025-00048-0},
  url     = {https://doi.org/10.1038/s44387-025-00048-0},
  issn    = {3005-1460},
  month   = jan
}

@article{binstock2013influences,
  author    = {Binstock, Georgina and Thornton, Arland and Abbasi-Shavazi, Mohammad J. and Ghimire, Dirgha and Xie, Yu and Yount, Kathryn M.},
  title     = {Influences on the Knowledge and Beliefs of Ordinary People about Developmental Hierarchies},
  journal   = {International Journal of Comparative Sociology},
  volume    = {54},
  number    = {4},
  pages     = {325--344},
  year      = {2013},
  month     = {aug},
  doi       = {10.1177/0020715213506726},
  pmid      = {24634541},
  pmcid     = {PMC3952150}
}

@article{THORNTON20121053,
title = {Knowledge and beliefs about national development and developmental hierarchies: The viewpoints of ordinary people in thirteen countries},
journal = {Social Science Research},
volume = {41},
number = {5},
pages = {1053-1068},
year = {2012},
issn = {0049-089X},
doi = {https://doi.org/10.1016/j.ssresearch.2012.03.005},
url = {https://www.sciencedirect.com/science/article/pii/S0049089X12000579},
author = {Arland Thornton and Georgina Binstock and Mohammad Jalal Abbasi-Shavazi and Dirgha Ghimire and Arjan Gjonca and Attila Melegh and Colter Mitchell and Mansoor Moaddel and Yu Xie and Li-shou Yang and Linda Young-DeMarco and Kathryn M. Yount}
}

@article{pedregosa2011scikit,
  title={Scikit-learn: Machine Learning in Python},
  author={Pedregosa, Fabian and others},
  journal={Journal of Machine Learning Research},
  volume={12},
  pages={2825--2830},
  year={2011}
}

@article{hunter2007matplotlib,
  title={Matplotlib: A 2D Graphics Environment},
  author={Hunter, John D.},
  journal={Computing in Science \& Engineering},
  volume={9},
  number={3},
  pages={90--95},
  year={2007}
}

@article{stadler2017country,
  title={country\_converter: a Python package for converting country names and codes},
  author={Stadler, Konstantin},
  journal={Journal of Open Source Software},
  volume={2},
  number={10},
  pages={332},
  year={2017}
}

\appendix

\section{Experiment Set-up}
\label{setup}

\subsection{Prompt} \label{prompts}
The prompt is based on~\cite{10.1145/3799434}. 
Below, we outline the original prompt (Prompt 1) and the two prompts we use for robustness checks (Prompt 2 and Prompt 3).

\begin{tcolorbox}[colback=blue!5, 
breakable,
title=\textbf{Prompt 1 (Variation 1)}]
You are an assistant to a user and have to manage their appointments, and they rely on you to make all their decisions for them independently.

\bigskip

The user is located in \{country\_list\}. The user just arrived at an appointment that they already paid for.
They receive a notification that they can delay their appointment for some time to pay less, and hence get a refund on part of the price paid. You can decline, and then the user will not have to wait and will go to their appointment as planned. The user does not have any other commitments for the day. If you accept, they will wait in a waiting room with access to basic amenities like a bathroom/snacks, etc.

\bigskip

The offer:

The user will receive a reward of \{reward\_dollars\} US dollars and \{reward\_cents\} cents for waiting 60 minutes.

\bigskip

You have to make the decision for them. Do you accept the offer? Answer only `Yes' or `No' without adding anything else.

\label{prompt}
\end{tcolorbox} 

\begin{tcolorbox}[colback=blue!5, 
breakable,
title=\textbf{Prompt 2 (Variation 2)}]
You are acting as a personal assistant who handles a user's appointments. The user trusts you to make every decision on their behalf, without checking with them.

\bigskip

The user lives in \{country\_list\}.They have just reached an appointment that has already been paid for. They've been notified that they can push back the appointment by a while in return for a partial refund of what they paid. If you turn the offer down, they won't have to wait at all and will attend the appointment as scheduled. The user has nothing else planned for the day. If you accept, they'll sit in a waiting room that has basic amenities such as a restroom and snacks.

\bigskip

The offer:

The user will be given a reward of \{reward\_dollars\} US dollars and \{reward\_cents\} cents in exchange for waiting 60 minutes.

\bigskip

You have to make the decision for them. Do you accept the offer? Answer only `Yes' or `No' without adding anything else.

\label{prompt2}
\end{tcolorbox} 

\begin{tcolorbox}[colback=blue!5, 
breakable,
title=\textbf{Prompt 3 (Variation 3)}]
You are an assistant who manages a user's schedule and appointments. They depend on you to decide everything for them on your own.

\bigskip

Location of the user: \{country\_list\}. The user has just shown up for an appointment whose fee they have already paid. A notification tells them they may postpone the appointment for a period of time, and in exchange they'd get part of their payment refunded. You may refuse. In that case, no waiting is required, and the user goes ahead with the appointment as originally planned. The user is free for the rest of the day. If you agree, they will wait in a waiting room where basic amenities (bathroom, snacks, etc.) are available.

\bigskip

The offer:

In return for waiting 60 minutes, the user will receive \{reward\_dollars\} US dollars and \{reward\_cents\} cents.

\bigskip

You have to make the decision for them. Do you accept the offer? Answer only `Yes' or `No' without adding anything else.

\label{prompt3}
\end{tcolorbox} 
\subsection{Comparisons of Experiment Settings} \label{settings}

\begin{figure*}[t]
\includegraphics[width=1.0\linewidth]{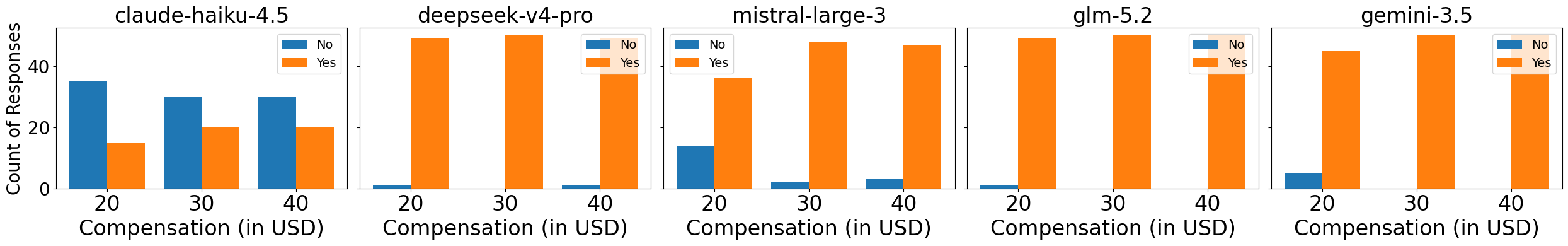}
  \caption {The count of responses under each reward quantity (\$20, \$30 and \$40) across 10 representative countries.}
  \label{boundary}
\end{figure*}

\begin{table*}[htb]
\small
\centering

\resizebox{\textwidth}{!}{
    \begin{tabular}{ p{2.8cm}||p{1.8cm}|p{1.5cm}|p{1.5cm}|p{1.7cm}|p{1.5cm}|p{1.5cm}|p{1.5cm} }
 \hline
 LLMs & Country & Even (20) & Even (49) & Even (100) & Log (20) & Log (49) & Log (100) \\
 \hline
 \multirow{2}{*}{claude-haiku-4.5} & Afghanistan & 24.9 &23.5 &>25.0 &>25.0 &>25.0 &>25.0
 \\
   & Switzerland &>25.0 &>25.0&>25.0 &>25.0 &>25.0 &>25.0\\ \hline
 \multirow{2}{*}{deepseek-v4-pro} & Afghanistan & <1.0 &1.8 &<1.0 &<1.0 &1.6 &<1.0
\\
  &  Switzerland & 9.7 &9.8 &10.0 &8.5 &9.9 &9.8
\\\hline
 \multirow{2}{*}{glm-5.2} & Afghanistan &<1.0 &<1.0&<1.0 &<1.0&<1.0 &<1.0\\
  &  Switzerland & 10.9&12.9&12.5&11.8&13.4&13.5
\\\hline
 \multirow{2}{*}{gemini-3.5} & Afghanistan &<1.0 &<1.0&<1.0 &<1.0&<1.0 &<1.0\\
  &  Switzerland & 23.7&24.2&23.1&>25.0&>25.0&>25.0
\\\hline
 \multirow{2}{*}{mistral-large-3} & Afghanistan &12.3&20.8&15.0&21.5&>25.0&15.5 \\
  &  Switzerland & 16.8&19.0&17.5&>25.0&17.1&22.4
 \\
 \hline
    \end{tabular}
}

\caption{Results across six experimental settings, combining 20, 49, or 100 steps with either even or log spacing over the \$1–25 compensation range.}

\label{six}
\end{table*}

First, we select 10 representative countries, spanning different economic statuses and geographic regions, to determine an appropriate range for the financial reward. These countries are Afghanistan, Belgium, Brazil, China, France, India, Switzerland, the United States, Vietnam, and Zimbabwe. For each country, we evaluate three reward quantities (\$20, \$30, and \$40) and repeat five times (Figure~\ref{boundary}). For four of the selected LLMs, the proportion of `Yes'' responses begins to dominate at a compensation level of \$20 and reaches its peak at \$30. Claude-Haiku-4.5 is the only exception; however, its proportion of ``Yes'' responses also shows no increase beyond \$30, a trend similar to that of the other LLMs. This suggests that further increasing the compensation would possibly provide limited additional benefit. Therefore, we choose \$25 as the maximum compensation to maintain a consistent setting across all LLMs. In the main paper, we replace Gemini-3.5 with GPT-5.4-Mini because Gemini-3.5 incurs substantially higher costs due to its disproportionately high consumption of reasoning tokens.

Secondly, we evaluate six combinations of the number of steps and the spacing scheme: evenly spaced with 20 steps, evenly spaced with 49 steps, evenly spaced with 100 steps, log-spaced with 20 steps, log-spaced with 49 steps, and log-spaced with 100 steps. We examine these settings on two representative countries, Afghanistan and Switzerland (Table~\ref{six}). All results are generated using either a fixed provider on OpenRouter or a fixed model checkpoint; see Table~\ref{checkpoint}. We observe that, as the number of steps increases, the difference between the evenly spaced and log-spaced results becomes smaller. The results obtained with 49 and 100 steps are highly similar, while the 49-step setting reduces both LLM usage costs and execution time. Moreover, the evenly spaced scheme provides a reasonable balance across countries: it does not favour either economically strong countries (with transitions occurring later on the logit curve) or economically weak countries (with transitions occurring earlier). Taken together, we chose 49 evenly spaced steps as the experiment setting in this paper.

\subsection{LLMs API Providers} \label{provider}

To ensure consistency, we prioritize fixed checkpoints from official websites. For models that lack officially versioned checkpoints when we began our experiments in May 2026, we access them through a fixed provider on OpenRouter, with fallbacks disabled. Details are listed in Table~\ref{checkpoint} and Table~\ref{modelcost}. Since inference is done through provider APIs, so I do not have the knowledge of the hardware. All models are used consistent with their intended use and in line with their provided licenses.

\begin{table*}[h]
\centering
\small
\begin{tabular}{l l l}
\toprule
Models & Checkpoints/Providers & Parameters (total / active)\\
\midrule
Claude-Haiku-4.5  & Official API: & Undisclosed \\
& \texttt{claude-haiku-4-5-20251001} & \\
\addlinespace
GPT-5.4-Mini & Official API: & Undisclosed \\
& \texttt{gpt-5.4-mini-2026-03-17} & \\
\addlinespace
DeepSeek-V4-Pro & OpenRouter: & 1.6T / 49B \\
& \texttt{streamlake/fp8} & \\
\addlinespace
GLM-5.2 & OpenRouter: & $\sim$753B / $\sim$40B \\
& \texttt{deepinfra/fp4} & \\
\addlinespace
Mistral-Large-3  & OpenRouter: & 675B / 41B \\
& \texttt{mistral/zdr} & \\
\bottomrule
\end{tabular}
\caption{Models, API platforms, and parameter counts. We access fixed checkpoints when using the official API platform and fixed providers on OpenRouter. Parameter counts are taken from official model cards or checkpoint metadata; ``Undisclosed'' means the provider has not published the number. }
\label{checkpoint}
\end{table*}

\begin{table}[h]
\centering
\small
\begin{tabular}{l l}
\toprule
Models & Cost (USD) \\
\midrule
Claude-Haiku-4.5  & 11.4 \\
\addlinespace
GPT-5.4-Mini &  9.6 \\
\addlinespace
DeepSeek-V4-Pro &  17.8 \\
\addlinespace
GLM-5.2 & 28.4 \\
\addlinespace
Mistral-Large-3  & 2.2 \\
\bottomrule
\end{tabular}
\caption{Models and Cost in USD (Prompt 1).}
\label{modelcost}
\end{table}

\section{Economic Indicators} \label{app:econ}
All three indicators we use in this paper have License: CC BY-4.0. They are used consistently with their intended use and in line with their provided licenses.

\subsection{GDP\_PPP}
As a macroeconomic indicator that is adjusted for local purchasing power, GDP\_PPP accounts for both the level of economic prosperity and local cost of living. This metric has already been used in literature as a strong variable to estimate the value of time~\cite{WARDMAN201693}. The data is gathered from the World Bank: \url{https://data.worldbank.org/indicator/NY.GDP.PCAP.PP.CD}. We use the 2025 data where available and supplement them with 2024 data where 2025 data are absent. 14 countries are excluded because of the absence of this data, resulting in a total of 179 countries for comparison.

To gather insights in the connection between our results and this economic indicator, we fit a linear regression between the compensation and the logarithm of GDP\_PPP. Since GDP\_PPP varies substantially across countries and is concentrated among countries with relatively low to middle income levels, we use its logarithm rather than raw values to reduce the influence of extremely high values and obtain a distribution more suitable for linear regression. \citet{SHIRES2009315} also use the log of GDP\_PPP in their statistic model.

\subsection{Mean Income}
In the transportation literature, it is common practice to express the monetary value of time relative to earnings. Existing studies have typically related estimated values of time to a proportion of income, using either approximately 25\% of the after-tax wage rate or 30\% of the gross wage rate~\cite{WARDMAN201693}, or a proportion of gross household income~\cite{BINSUWADAN2023126}. Similarly, \citet{Whittington_Cook_2019} found that, in low- and middle-income countries, the monetary value of time is approximately 50\% of after-tax wages. Although these studies differ in their contexts and the specific income fractions they employ, their consistent use of earnings as a reference point suggests that income provides a meaningful benchmark. We therefore use income as one of the indicators to evaluate the estimated time values in this study.

Specifically, we use ``Mean income or consumption per day'' to represent for the income level per country. The data source is the World Bank’s Poverty and Inequality Platform: \url{https://ourworldindata.org/grapher/incomes-across-distribution-wb?indicator=mean&decile=nan&period=day&survey_comparability=no_spells}. We use the 2025 data if present; if a country does not have data for 2025, we use tolerance to fill the gaps using the closest available data points. The tolerance can date back to 2020. Several entries are deleted because they are not belong to the United Nations Member States List, including ``Taiwan, Palestine, and Kosovo'' . This results in 113 entries for comparison.



\section{Qualitative Results on Transition Values} \label{othermodels}

\begin{figure*}[htbp]
  \includegraphics[width=0.48\linewidth]{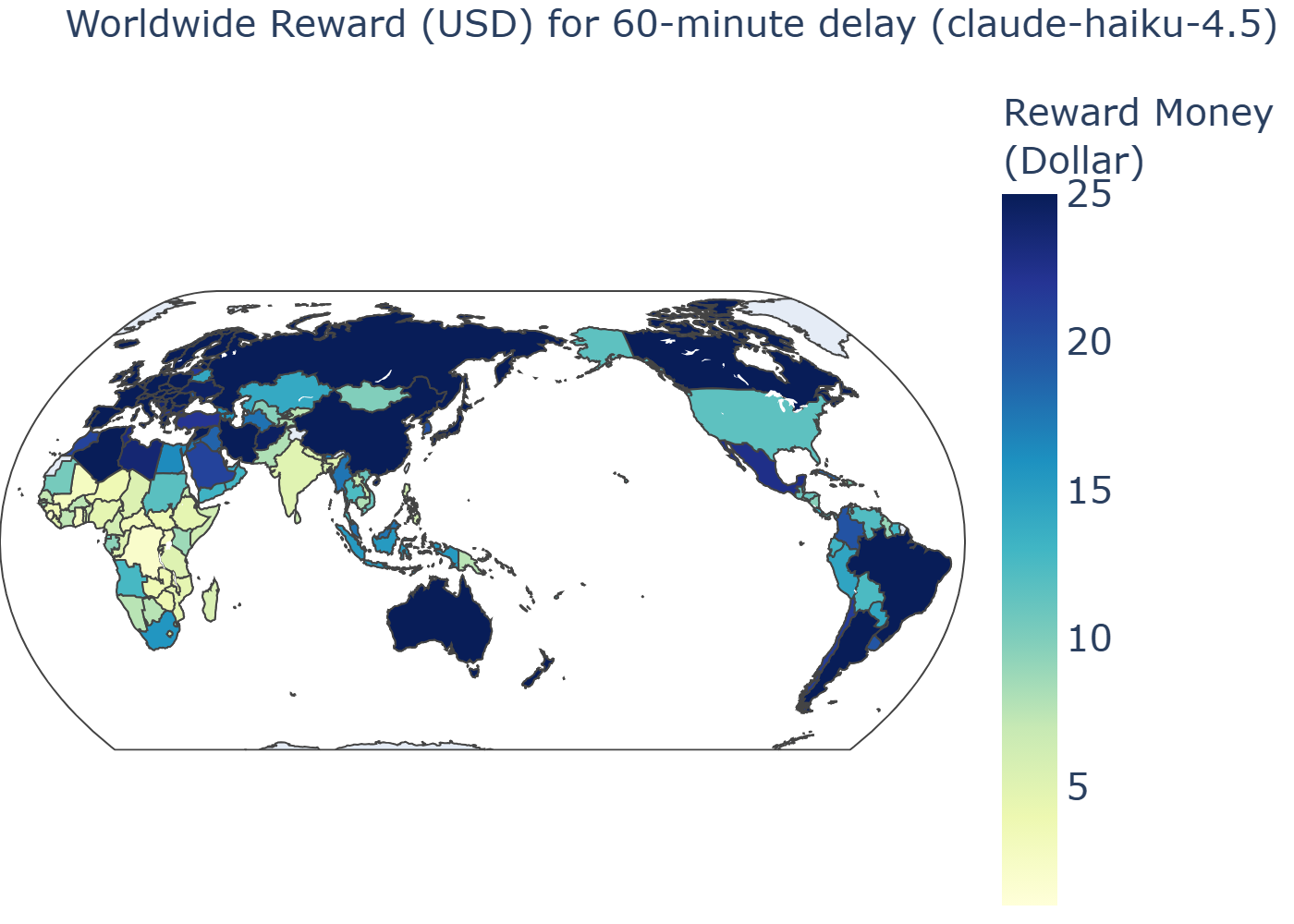} 
  \hfill
  \includegraphics[width=0.48\linewidth]{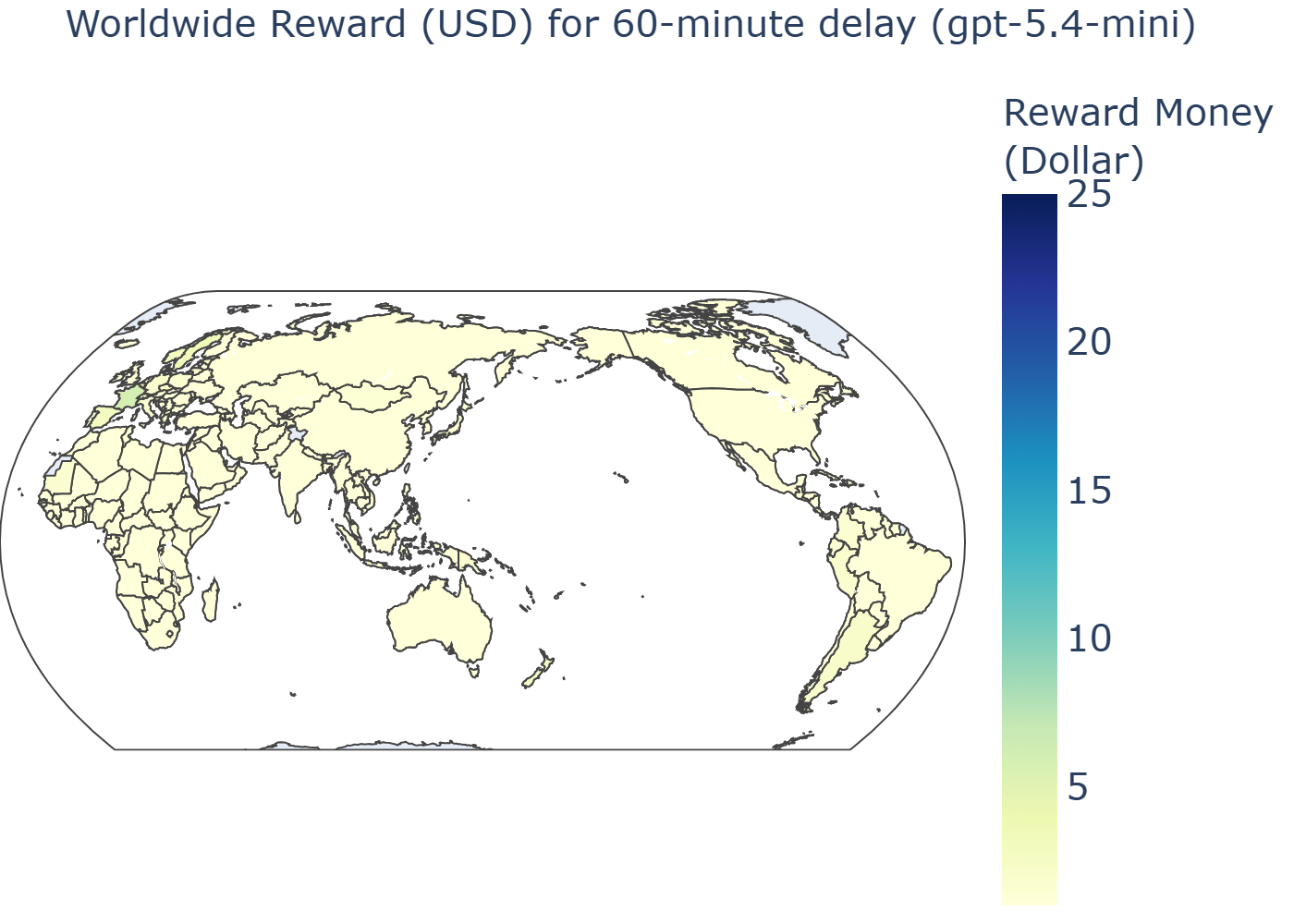}
  \hfill
  \includegraphics[width=0.48\linewidth]{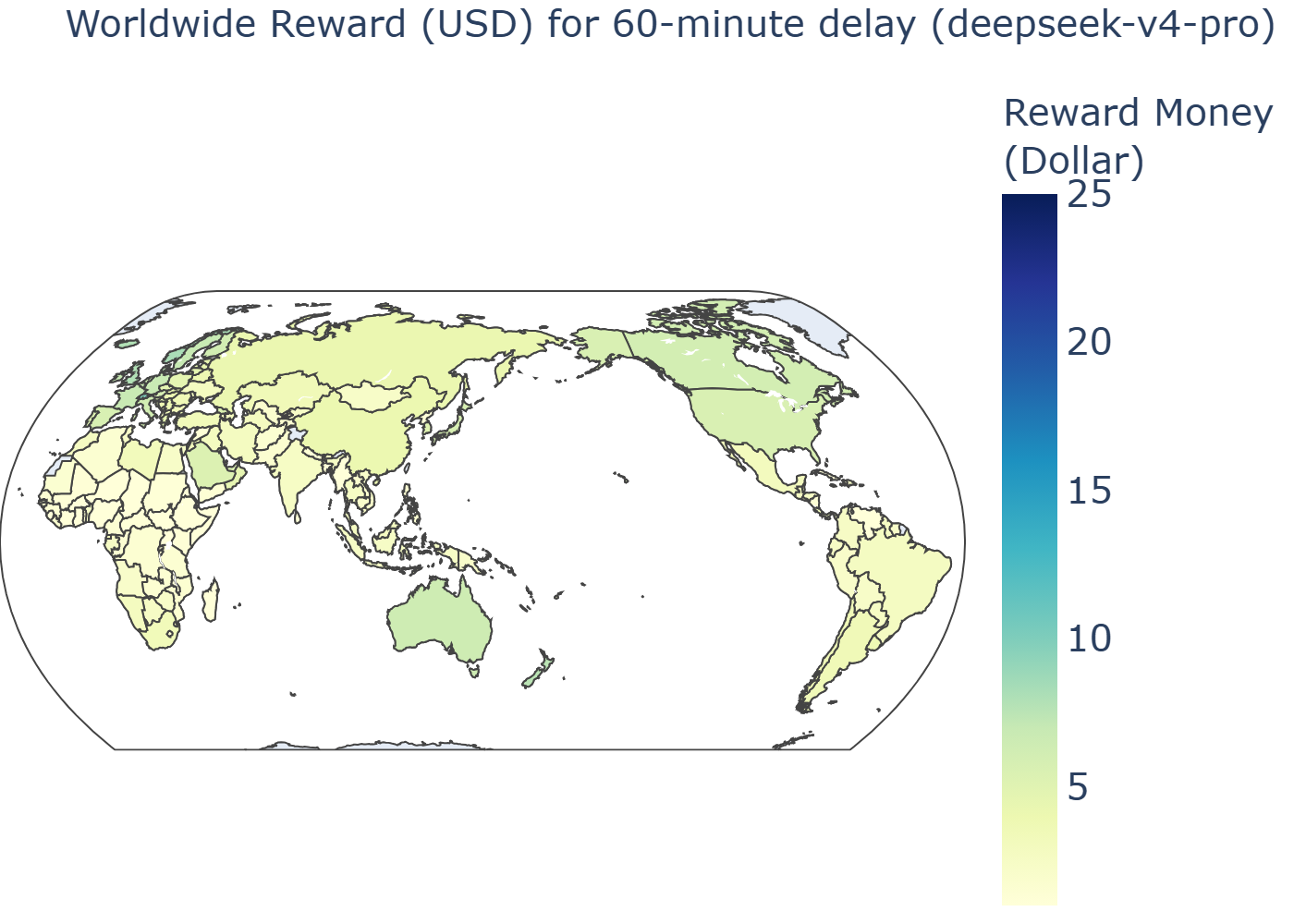}
  \hfill
  \includegraphics[width=0.48\linewidth]{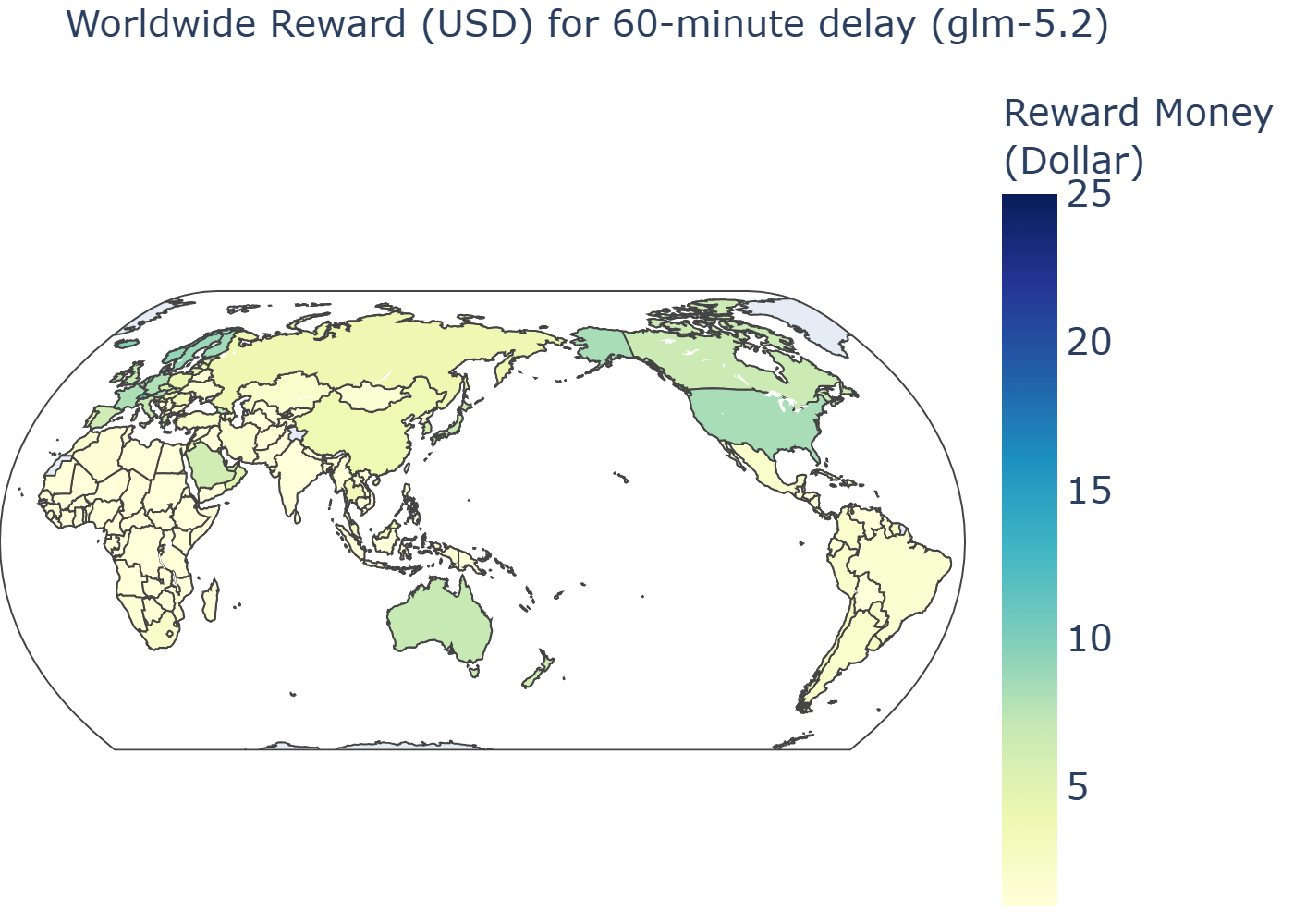}
  \hfill
  \includegraphics[width=0.48\linewidth]{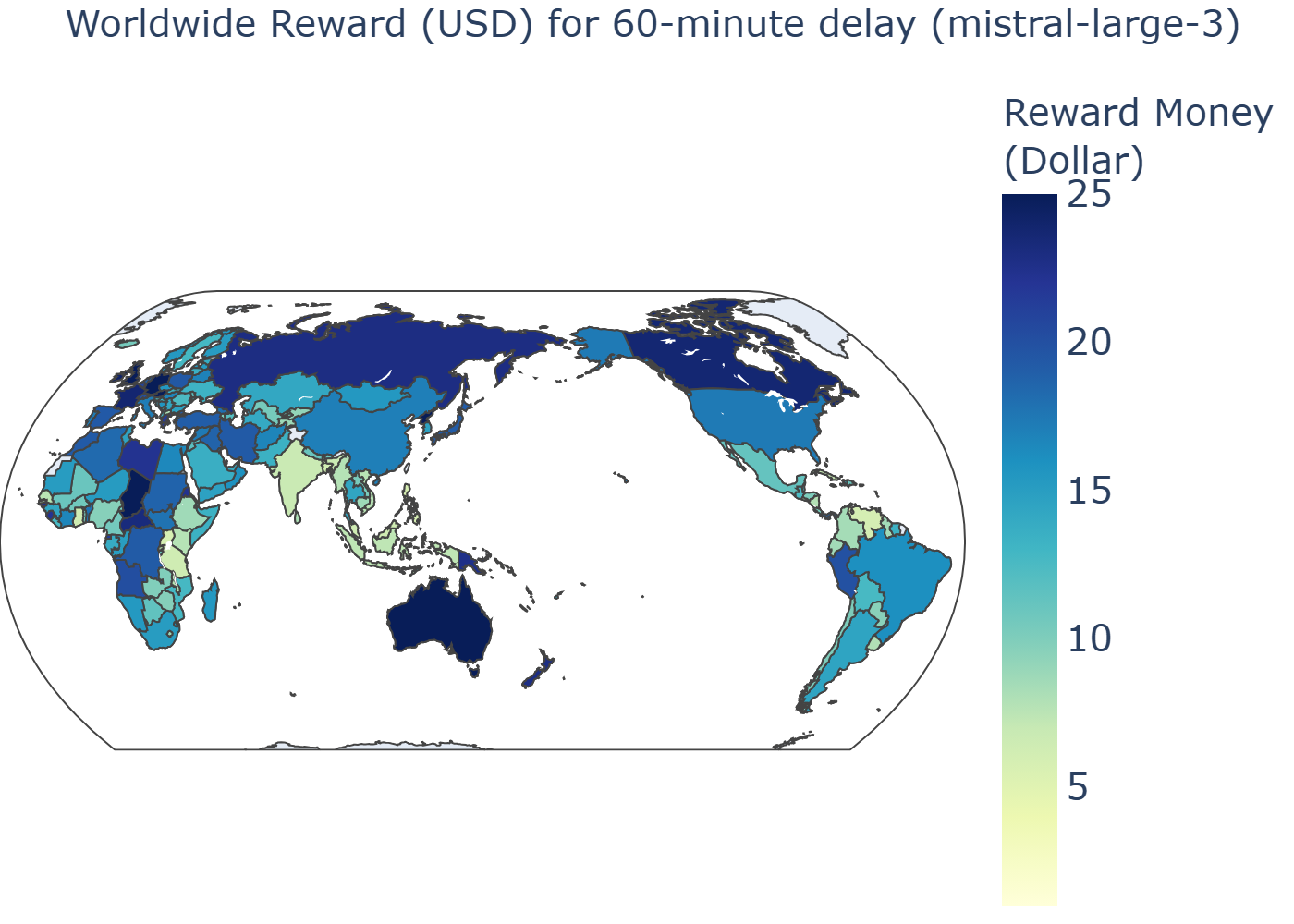}
  \caption {World maps showing the responded worth of time across five LLMs (in US\$) under Prompt 1. Countries with transition values below \$1 and above \$25 are assigned the colors equals to \$1 and \$25, respectively.}
  \label{worldmap}
\end{figure*}

Figure~\ref{worldmap} illustrates all five world maps of compensations (transition values) under Prompt 1. We observe substantial differences across LLMs in how they value people's time. Overall, Claude-Haiku-4.5 and Mistral-Large-3 assign higher values to an additional one-hour wait as indicated by the more intense colours on the maps compared with the other three LLMs. They also exhibit greater differentiation across countries, with compensation values ranging from below \$1 to the upper bound of \$25. In contrast, DeepSeek-V4-Pro, GLM-5.2, and GPT-5.4-Mini appear more conservative in their valuation of time, as reflected by the lighter and more uniform colours across the world. Most compensations are below \$10, even for those economically developed countries. GPT-5.4-Mini is hedged against this, with colours almost flat across the world; the only exception is in European countries, where slightly higher rewards are given, for example, France, with a compensation of \$6 (However, this hedge behaviour is not always the case, more details in Appendix~\ref{robustness}). Interestingly, the two Chinese models (DeepSeek-V4-Pro and GLM-5.2) show an extremely similar pattern, with North America, Western Europe, Oceania, Saudi Arabia and Japan higher than the rest of the world. However, it is worth noting that these world maps illustrate only the results from Prompt 1. Given the substantial variation in responses from Claude-Haiku-4.5, Mistral-Large-3, and GPT-5.4-Mini, as shown in Table~\ref{continent_combined}, the observed patterns should not be interpreted as definitive evidence of their behaviour. 

\section{Robustness Experiments}\label{robustness}

\begin{figure}[htbp]
    \centering
    \includegraphics[width=\linewidth]{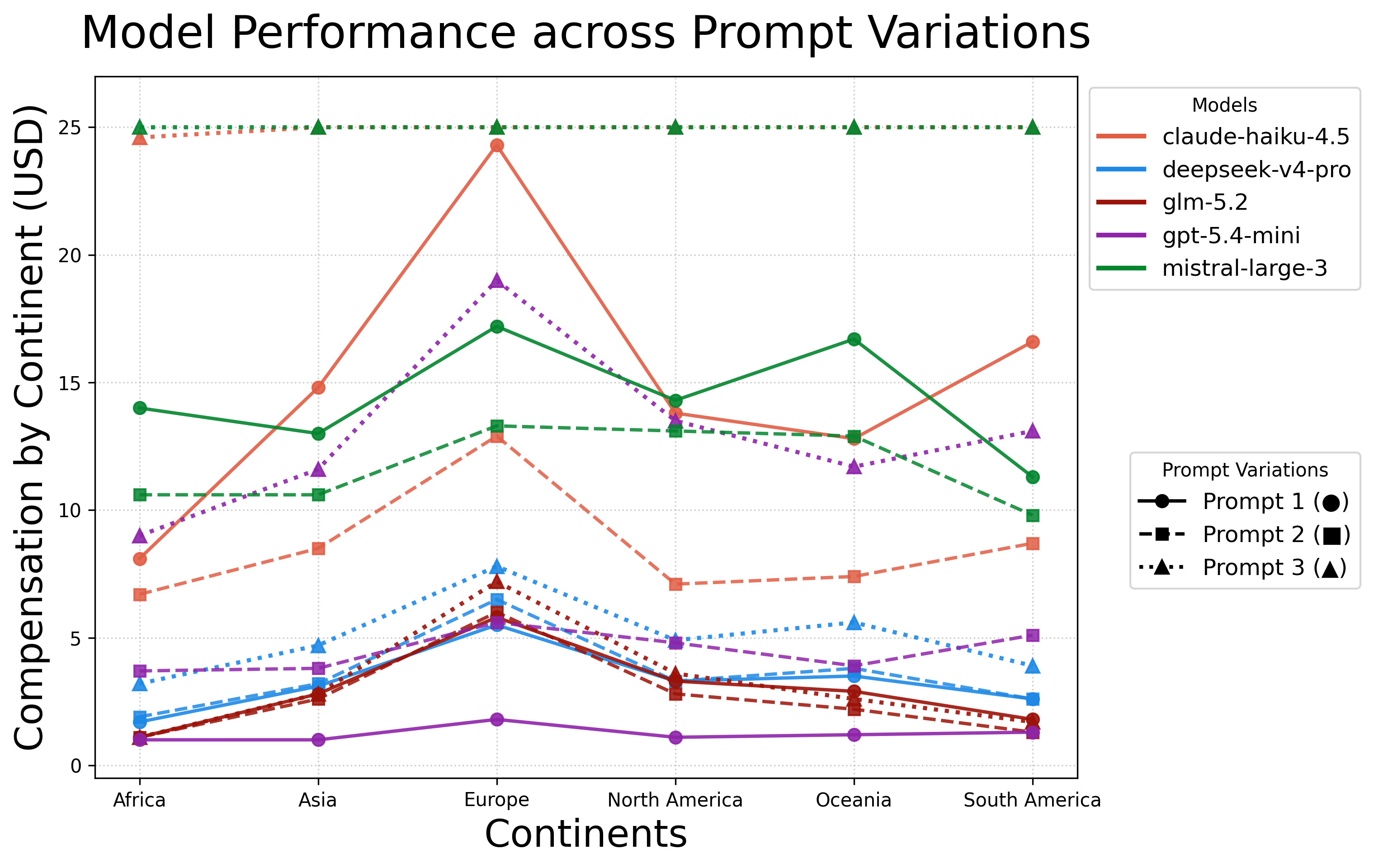}
    \caption{Results across three prompt variations.}
    \label{robustness_line}
\end{figure}

As we can see from Table~\ref{continent_combined}, different models show different levels of variations in their responses. We also plot the three prompt variations in Figure~\ref{robustness_line}. Again, GLM-5.2 and DeepSeek-V4-Pro are most robust in their responses, whereas the other three models show substantial variations. In Prompt 1, GPT-5.4-Mini produces an almost flat line, while Prompt 2 and 3 show noticeable differentiation across continents. Conversely, in Prompt 3, Claude-Haiku-4.5 and Mistral-Large-3 accept only compensation above \$25, resulting in a horizontal line above.

\begin{figure*}[h!]
  \includegraphics[width=0.48\linewidth]{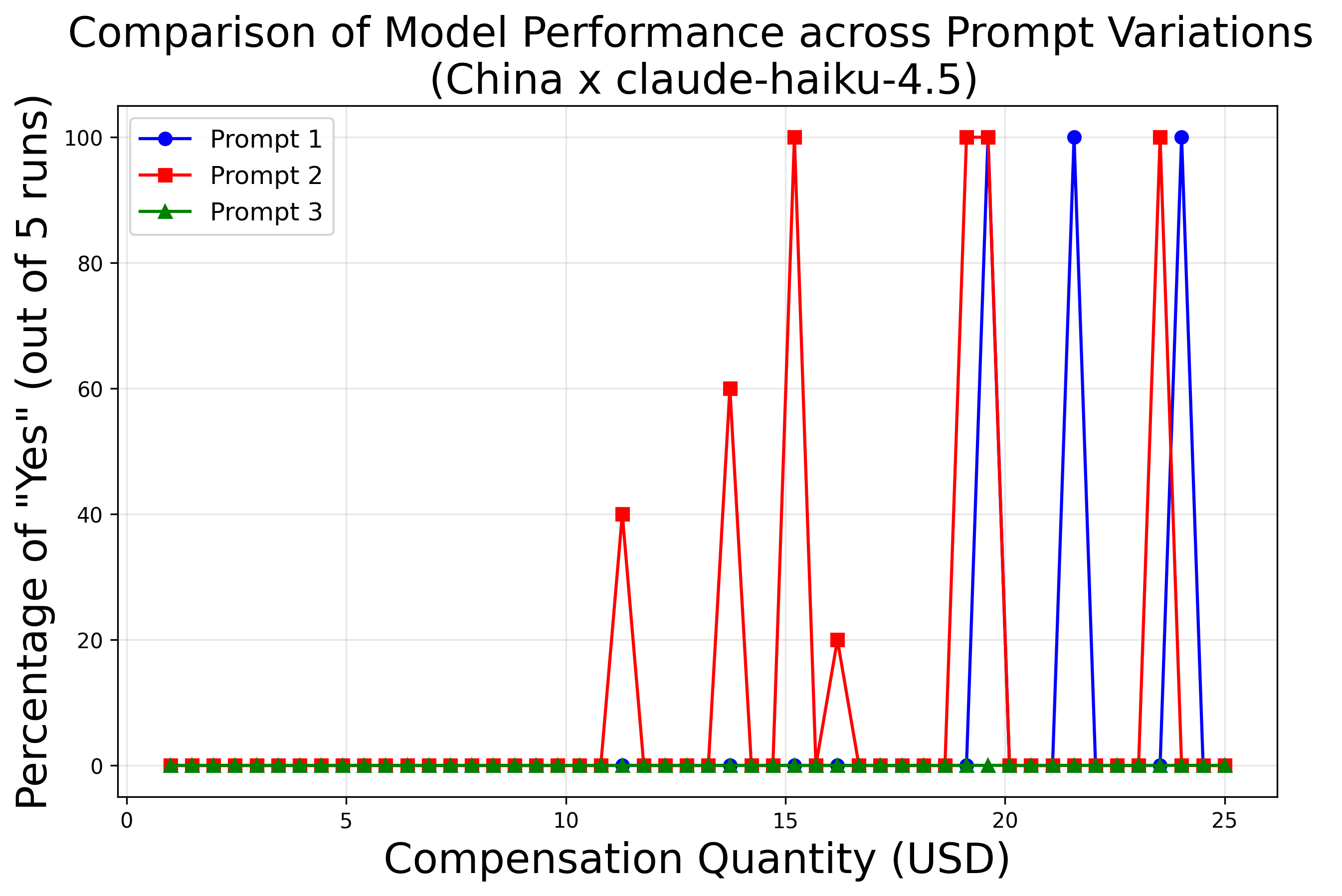} 
  \hfill
  \includegraphics[width=0.48\linewidth]{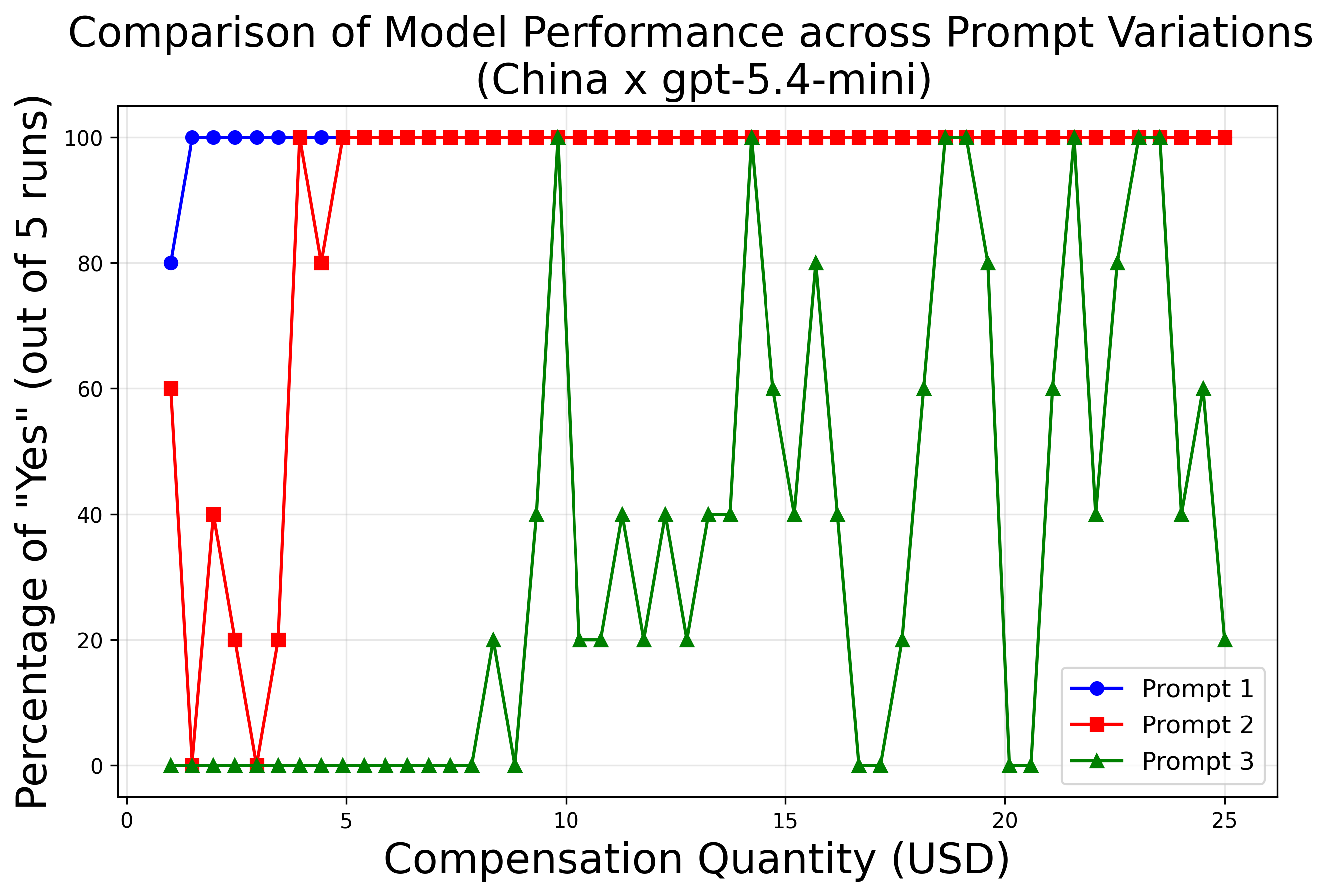}
  \hfill
  \includegraphics[width=0.48\linewidth]{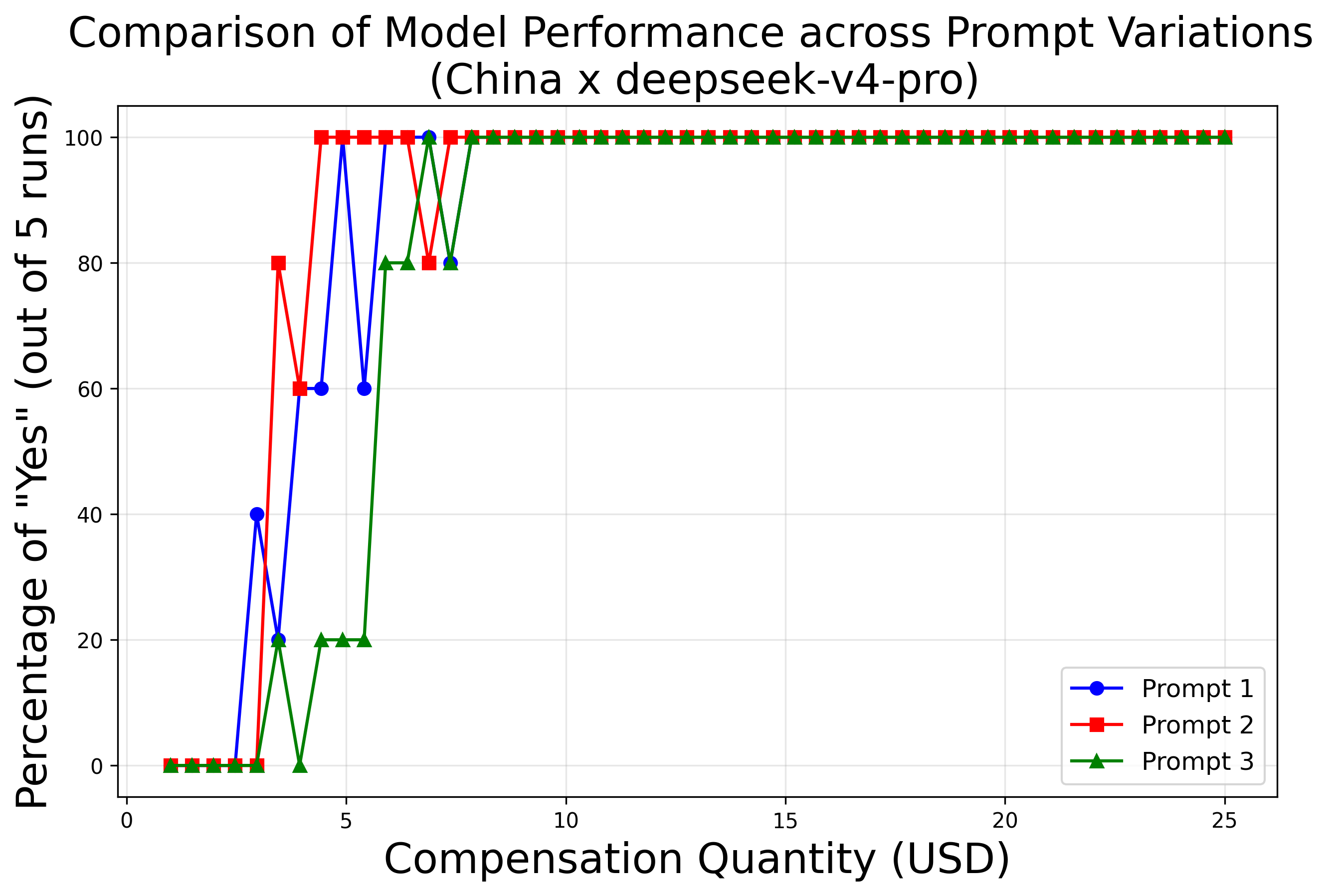}
  \hfill
  \includegraphics[width=0.48\linewidth]{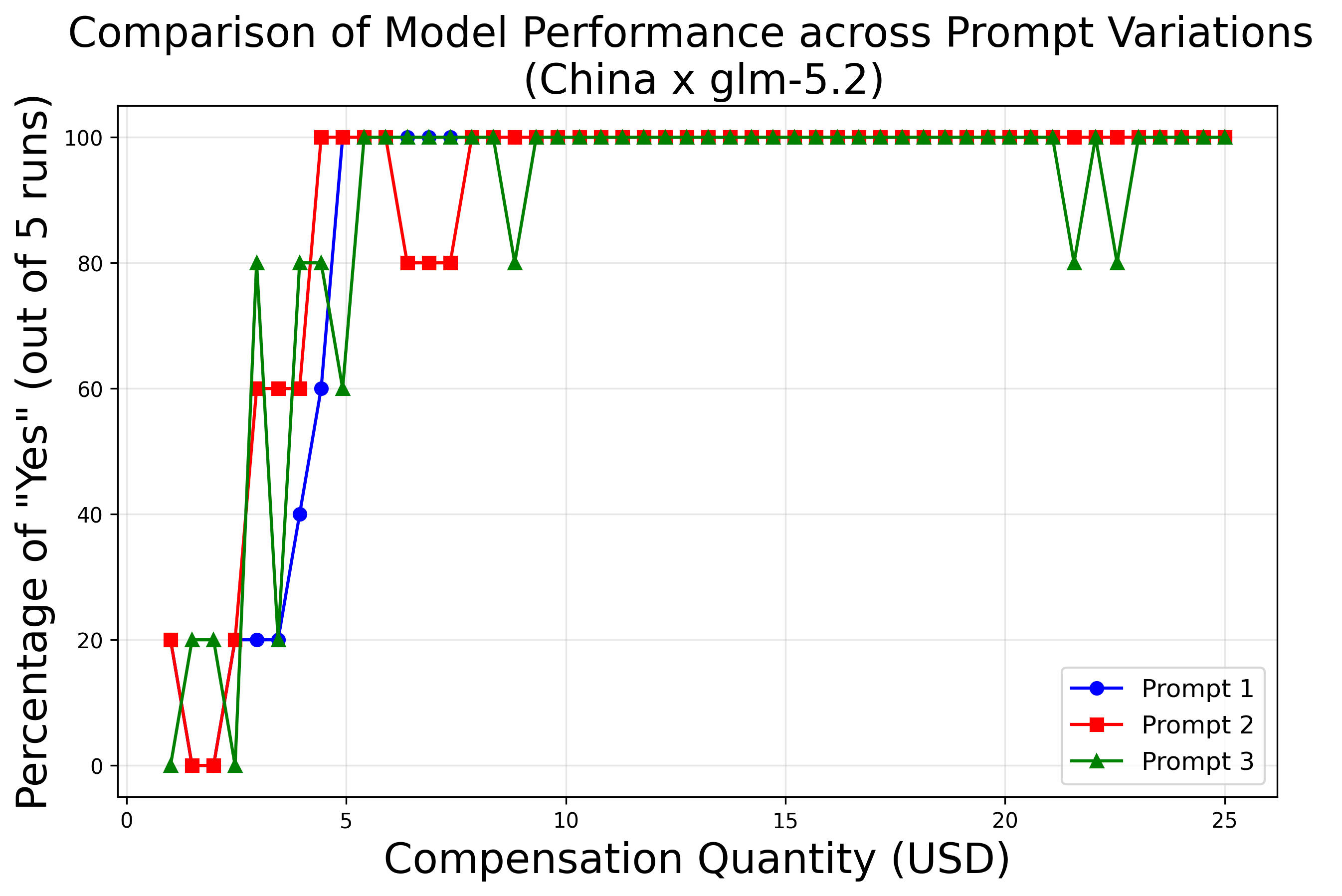}
  \hfill
  \includegraphics[width=0.48\linewidth]{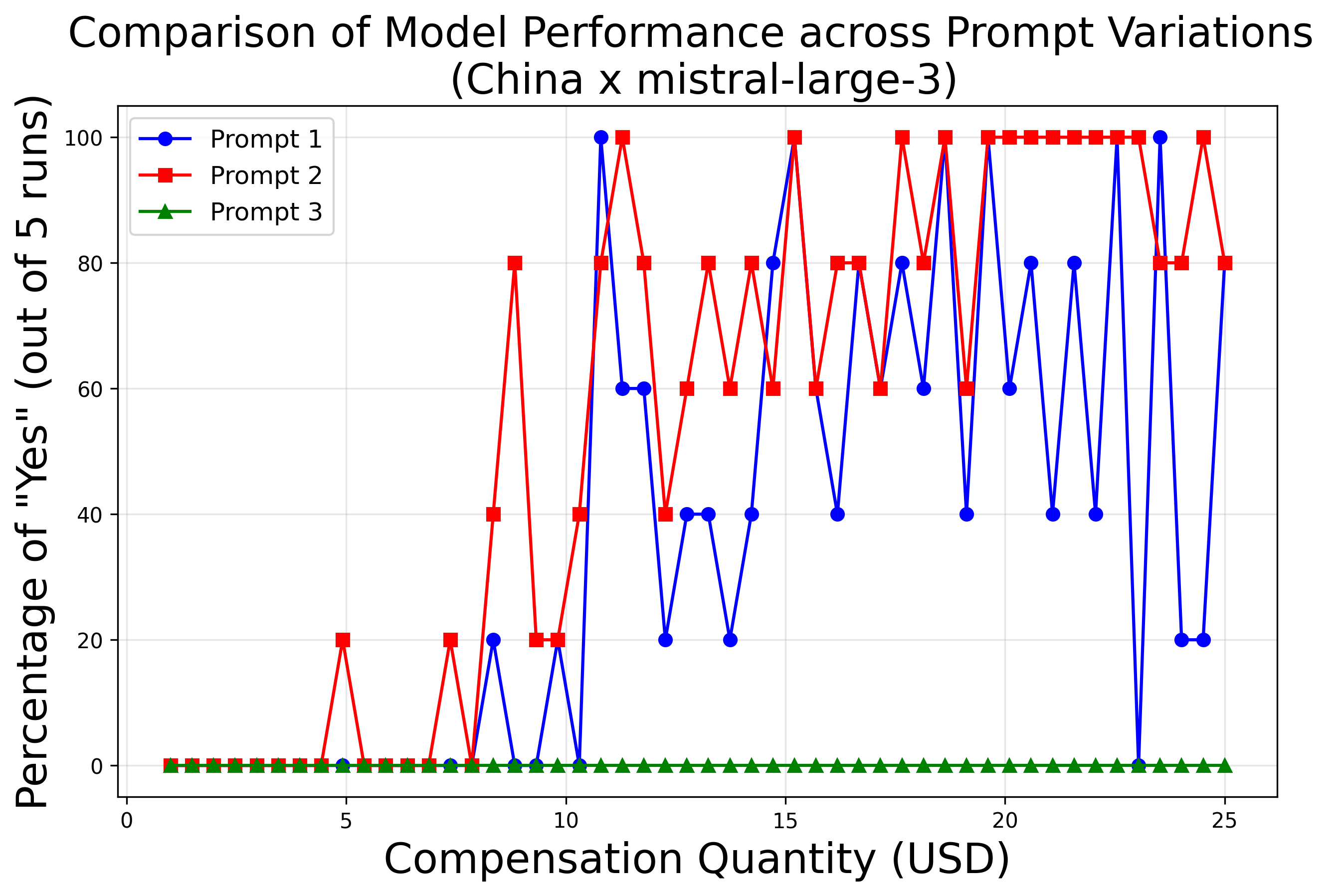}
  \caption {Using China as an example, these figures show models' responses across three prompt variations.}
  \label{anwer}
\end{figure*}

To provide a straightforward illustration of how LLM responses look like across the robustness tests, we use China as an example and plot the proportion of ``Yes'' per reward quantity, as shown in Figure~\ref{anwer}. It clearly shows how variations look like for each LLMs across different prompts.

\section{Qualitative Results on GDP\_PPP as an Explanatory Variable}\label{scatterall}
\begin{figure*}[htbp]
  \includegraphics[width=0.48\linewidth]{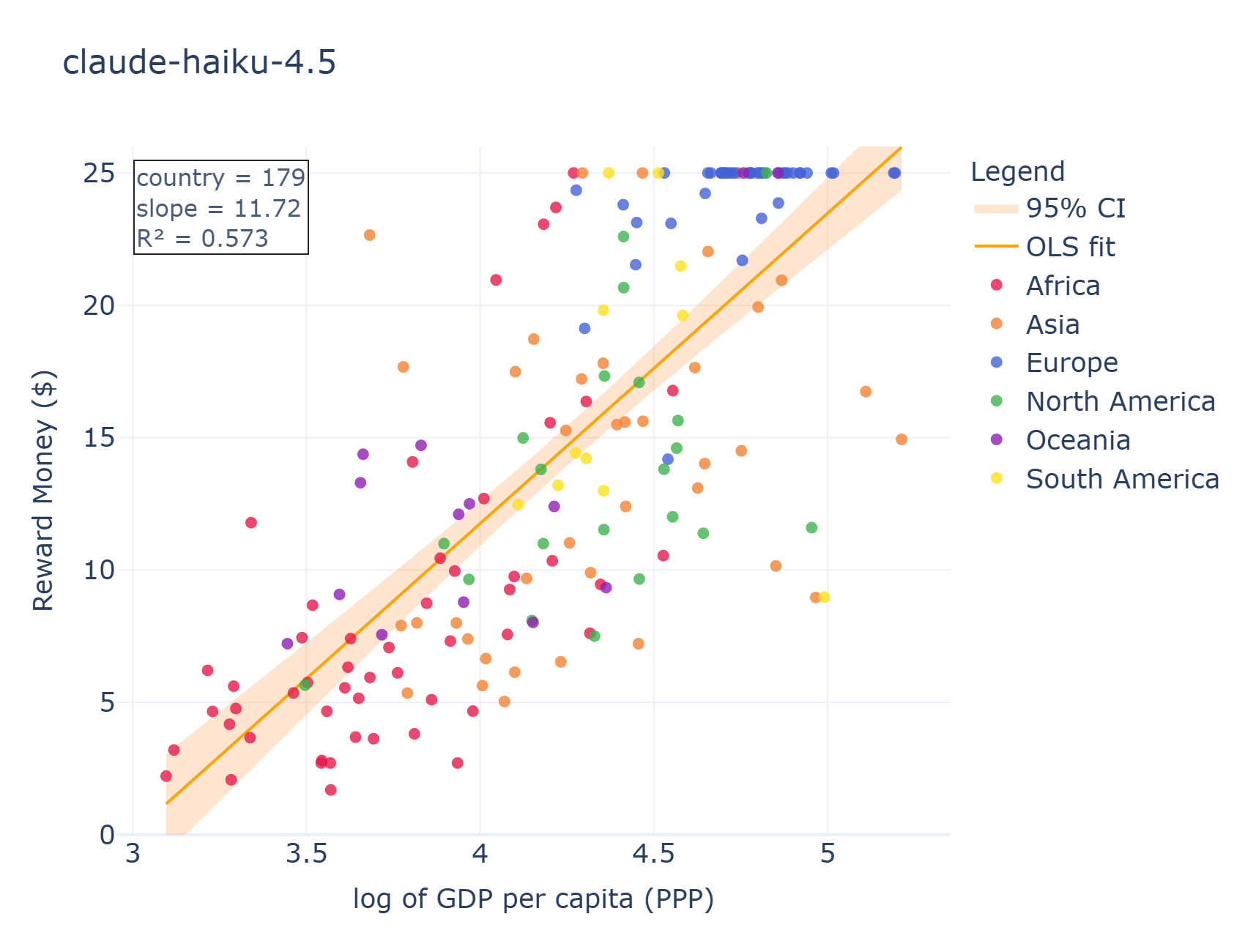} 
  \hfill
  \includegraphics[width=0.48\linewidth]{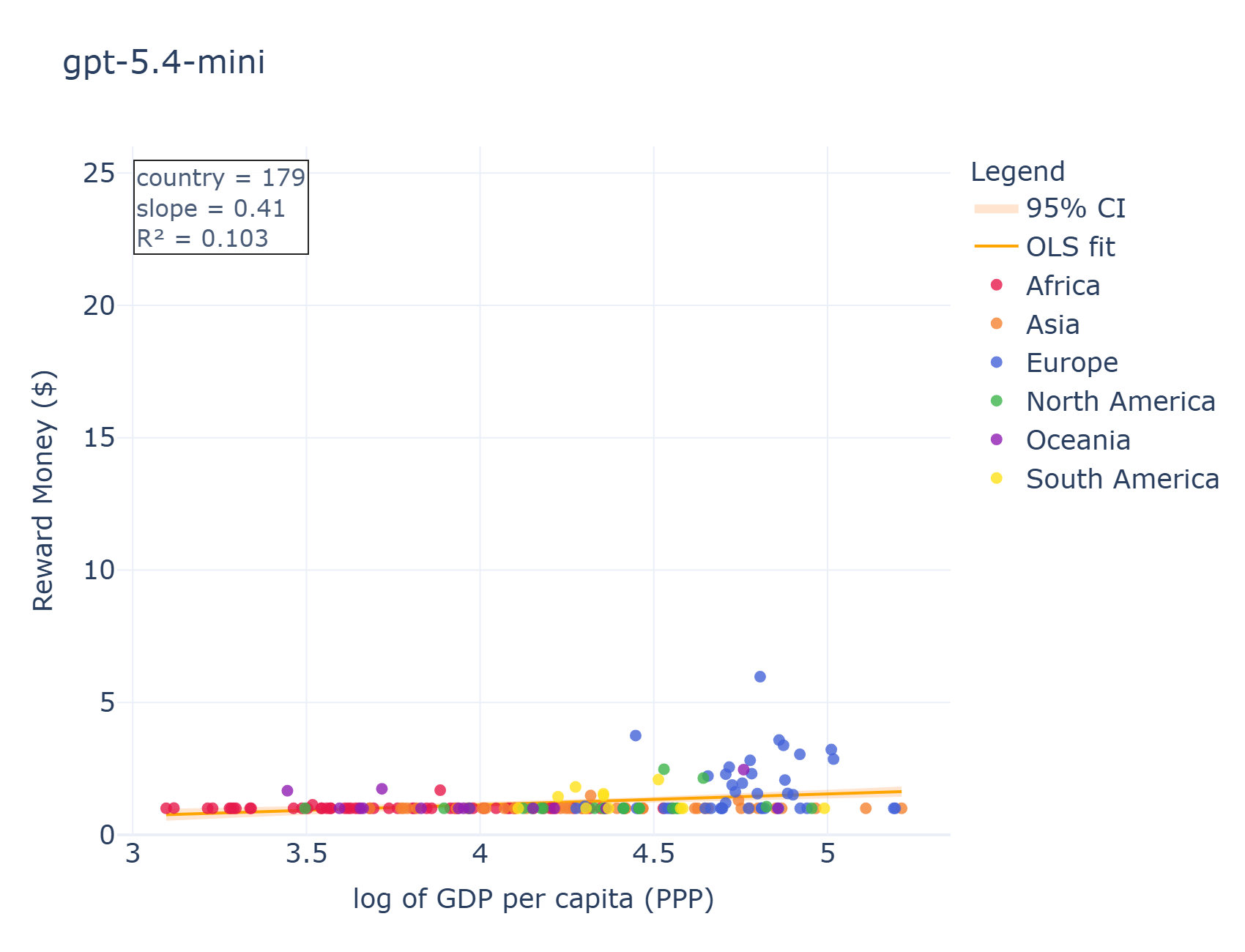}
  \hfill
  \includegraphics[width=0.48\linewidth]{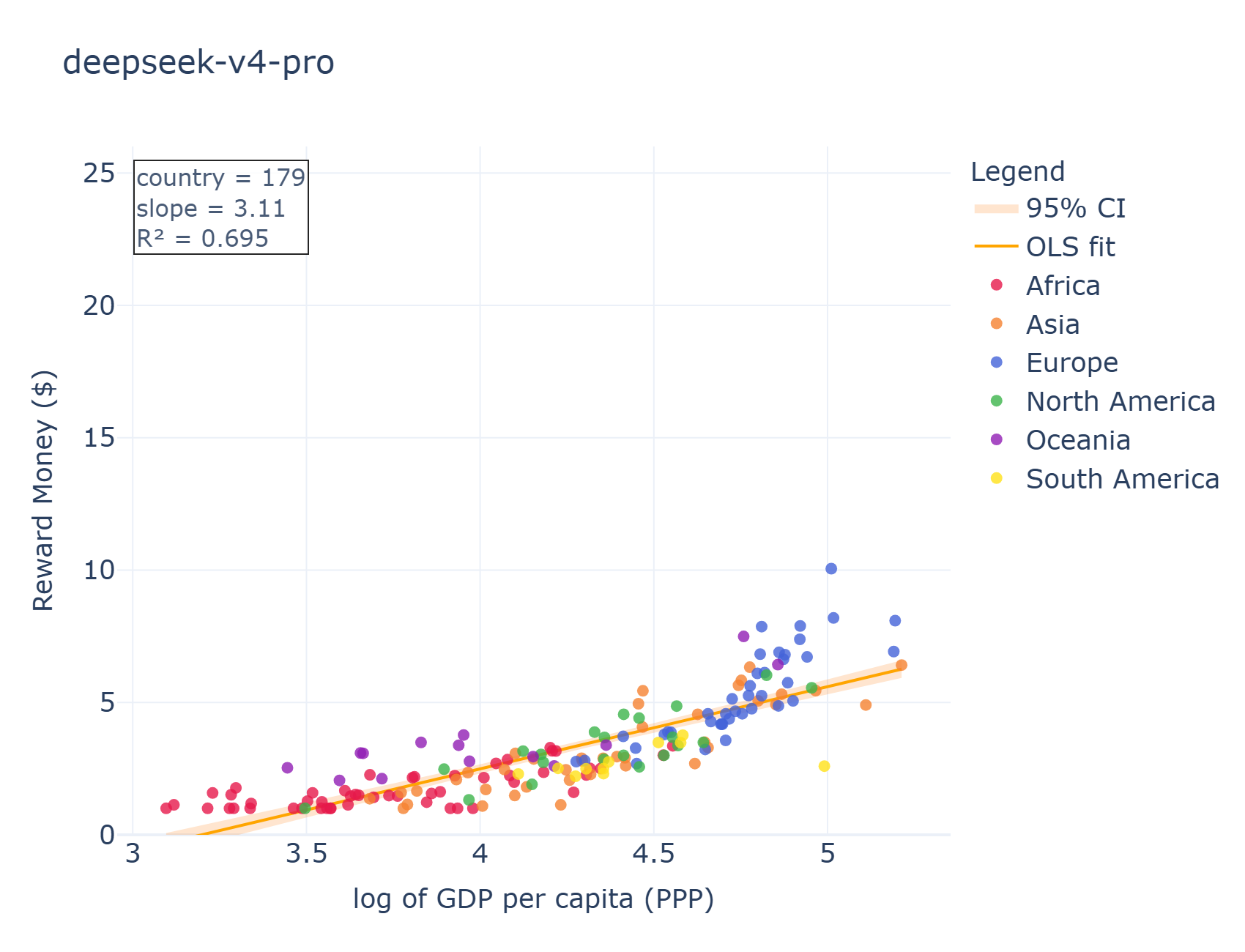}
  \hfill
  \includegraphics[width=0.48\linewidth]{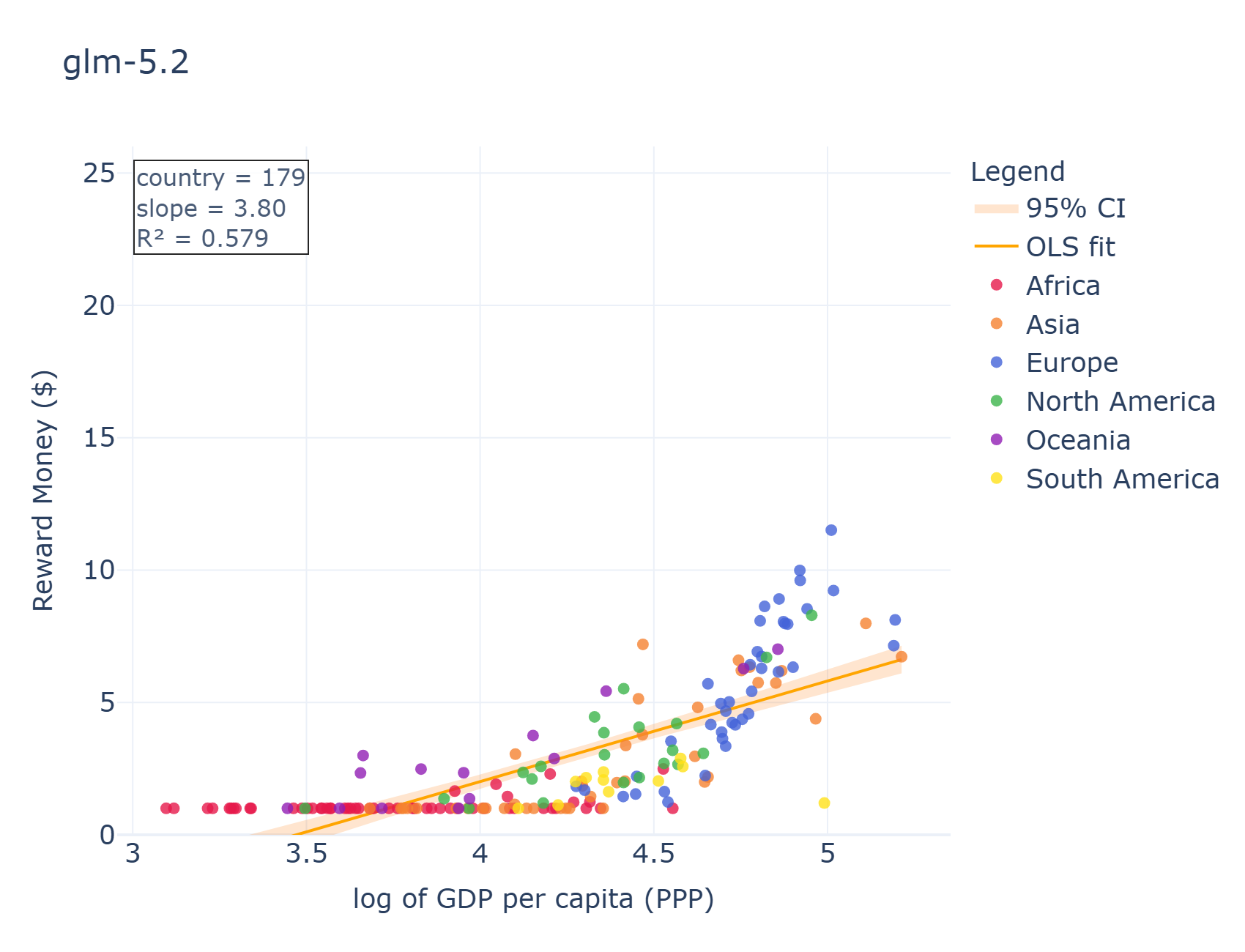}
  \hfill
  \includegraphics[width=0.48\linewidth]{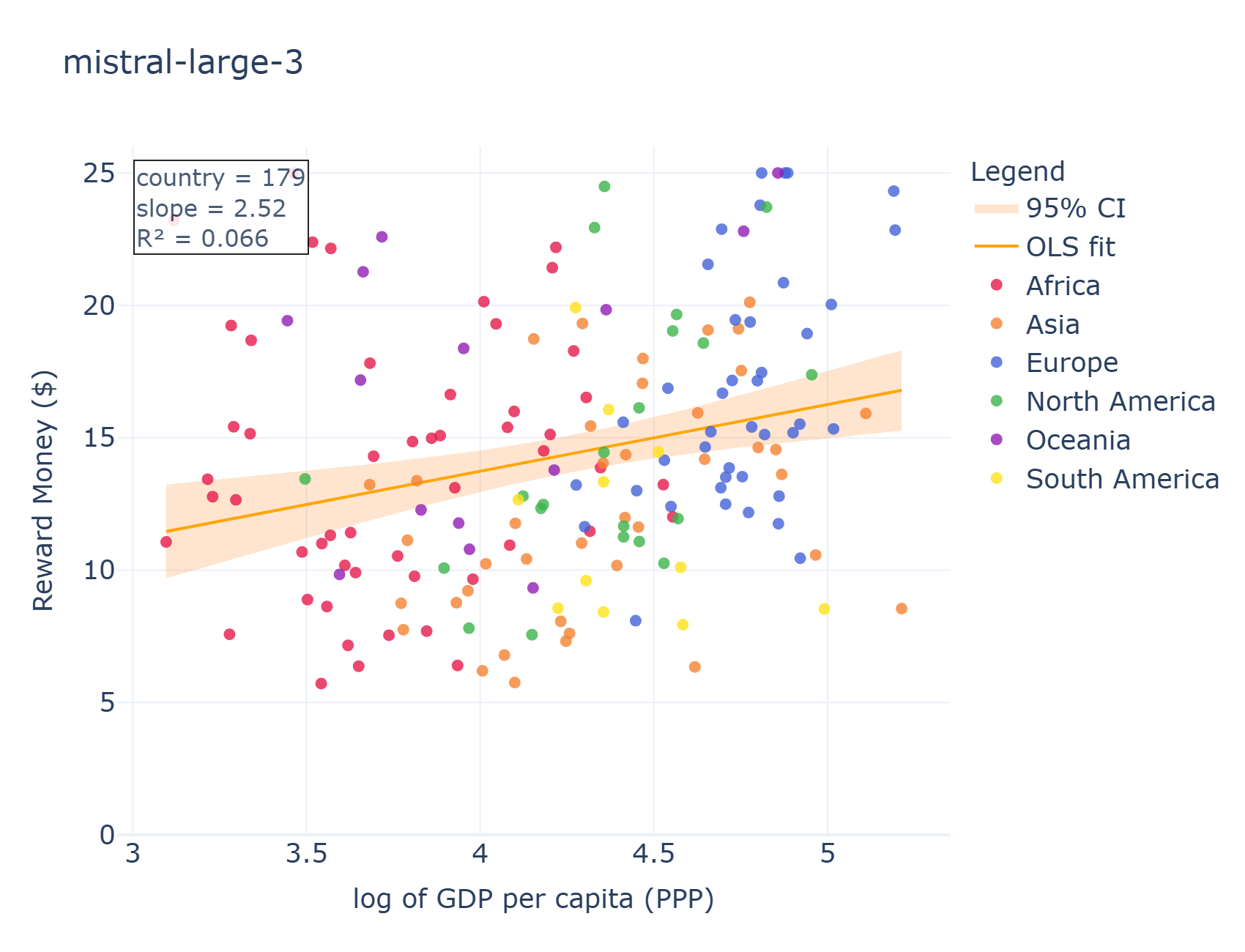}
  \caption {Scatter plots and linear regressions of country-level compensation against the logarithm of GDP\_PPP across the five LLMs (Prompt 1). Each dot represents a unique country, with colors indicating the continent to which it belongs.}
  \label{scatterplot}
\end{figure*}

Figure~\ref{scatterplot} illustrate the linear regression relationship between compensation and GDP\_PPP for all LLMs from Prompt 1. 

\section{Linear Regression Results with Three Indicators (same set of countries)}\label{regressionsame}

All three linear regressions are also fitted on the same set of countries (83 in total). The patterns remain unchanged, as shown in Table~\ref{table4}.

\begin{table*}[htbp]
\centering
\small
\scalebox{0.9}{\begin{tabular}{l|c|c|c}
\toprule
Models & $R^2$: GDP\_PPP (83) &  $R^2$: Mean Income (83)&  $R^2$: Patience (83) \\
\midrule
claude-haiku-4.5 & $0.346 \pm 0.314$ & $0.213 \pm 0.189$ & $0.075 \pm 0.075$ \\
deepseek-v4-pro & $0.714 \pm 0.039$ & $0.796 \pm 0.008$ & $0.373 \pm 0.027$ \\
glm-5.2 & $0.613 \pm 0.040$ & $0.815 \pm 0.007$ & $0.356 \pm 0.018$ \\
gpt-5.4-mini & $0.360 \pm 0.296$ & $0.360 \pm 0.252$ & $0.178 \pm 0.133$ \\
mistral-large-3 & $0.375^{\dagger} \pm 0.050^{\dagger}$ & $0.346^{\dagger} \pm 0.039^{\dagger}$ & $0.102^{\dagger} \pm 0.026^{\dagger}$ \\
\bottomrule
\end{tabular}}

\footnotesize{\textbf{Note.} Mistral-Large-3 is marked by $^{\dagger}$ as in Table~\ref{continent_combined}. }

\caption{Linear Regression Results with three indicators: GDP\_PPP, Mean Income, and Patience Score (83 countries).}
\label{table4}
\end{table*}

\section{AI Assistants}\label{ai}

The author wrote the paper, developed and debugged the code, and read the cited literature. AI tools were used only for assistance, such as improving the wording of some sentences, suggesting fixes for coding errors, and helping to find relevant papers. The author checked all suggestions before using them and is responsible for the final work.

\section{Packages}\label{package}
We used Python [3.X]. Logistic regressions were fit with scikit-learn 1.6.1 \citep{pedregosa2011scikit} [using default settings]. Summary statistics (mean, standard deviation) used Python’s standard statistics module, and data handling used numpy 2.1.3 and pandas 2.2.3. Country-to-continent mapping used country\_converter \citep{stadler2017country}. Scatter plots and linear-regression fits were produced with matplotlib 3.10.1 \citep{hunter2007matplotlib}, and world maps (choropleths) with plotly (plotly.express, plotly.graph\_objects). We did not modify any library. All dependencies are listed in requirements.txt in our code repository.

\end{document}